\documentstyle[doublespace,aps]{revtex}
\begin{document}
\draft
\author{DJ Wagner$^{1,2}$ and Thomas J. Weiler$^1$}
\address{$^1$Department of Physics and Astronomy, Vanderbilt University, Box 
   1807 B, Nashville, TN  37235}
\address{$^2$Department of Physics and Astronomy, Angelo State University, 
   Box 10904, San Angelo, TX  76909}
\date{\today}
\title{Boxing with Neutrino Oscillations}
\maketitle
\begin{abstract}
We develop a model-independent ``box" parameterization of neutrino oscillations.
Oscillation probabilities are linear in these new parameters, so  
measurements can straightforwardly determine the box parameters
which can then be manipulated to yield magnitudes of mixing matrix elements.
We examine the effects of unitarity 
on the box parameters and reduce the number of parameters to the minimum set. 
Using the box algebra, we show 
that CP-violation may be inferred from measurements
of neutrino flavor mixing even when the oscillatory factor has averaged.
The framework presented here will facilitate general analyses of neutrino
oscillations among $n\ge 3$ flavors.
\end{abstract}
%
\def\beq{\begin{equation}}
\def\eeq{\end{equation}}
\def\bea{\begin{eqnarray}}
\def\eea{\end{eqnarray}}
\def\ba{\begin{array}}
\def\ea{\end{array}}
\def\ab{{\alpha \beta}}
\def\acb{{\alpha, \beta}}
\def\gsim{\raisebox{-0.5ex}{$\stackrel{>}{\sim}$}}
\def\lsim{\raisebox{-0.5ex}{$\stackrel{<}{\sim}$}}
\def\half{{\frac{1}{2}}}
\def\openone{\leavevmode\!\!\!\!\hbox{\small1\kern-1.55ex\normalsize1}}
\def\nue{{\nu_{e}}}
\def\numu{{\nu_{\mu}}}
\def\nutau{{\nu_{\tau}}}
\def\nui{{\nu_{i}}}
\def\nua{{\nu_\alpha}}
\def\nub{{\nu_\beta}}
\def\Pab{{P\makebox[9 mm][r]{\raisebox{-1.5ex}{{\scriptsize{$\nua \rightarrow
\nub$}}}}} \!\!\!\!\!\!\!\!\!\!\!(x)\;}
\def\Paa{{P\makebox[9 mm][r]{\raisebox{-1.5ex}{{\scriptsize{$\nua \rightarrow
\nua$}}}}} \!\!\!\!\!\!\!\!\!\!\!(x)\;}
\def\Pba{{P\makebox[9 mm][r]{\raisebox{-1.5ex}{{\scriptsize{$\nub \rightarrow
\nua$}}}}} \!\!\!\!\!\!\!\!\!(x)\;}
\def\Peu{{P\makebox[8 mm][r]{\raisebox{-1.5ex}{{\scriptsize{$\nue \rightarrow
\numu$}}}}} \!\!\!\!\!\!\!\!\!(x)\;}
\def\Put{{P\makebox[8 mm][r]{\raisebox{-1.5ex}{{\scriptsize{$\numu \rightarrow
\nutau$}}}}} \!\!\!\!\!\!\!\!\!(x)\;}
\def\Pet{{P\makebox[8 mm][r]{\raisebox{-1.5ex}{{\scriptsize{$\nue \rightarrow
\nutau$}}}}} \!\!\!\!\!\!\!\!\!(x)\;}
\def\Pabbar{{P\makebox[9 mm][r]{\raisebox{-1.8ex}{{\scriptsize{
$\overline{\nu}_{alpha} \rightarrow \overline{\nu}_{\beta}$}}}}} 
\!\!\!\!\!\!\!\!\!(x)\;}
\def\Pbabar{{P\makebox[9 mm][r]{\raisebox{-1.8ex}{{\scriptsize{
$\overline{\nu\nu}_{\beta} \rightarrow \overline{\nu}_{alpha}$}}}}} 
\!\!\!\!\!\!\!\!\!(x)\;}
\newcommand{\Pnu}[2]{{P\!\!\!\!\makebox[9 mm][r]{\raisebox{-0.8ex}
  {{\tiny{${#1} \!\!\rightarrow \!\!{#2}$}}}}}}
\newcommand{\B}[2]{\;^{#1}\Box_{#2}\;}
\newcommand{\Bs}[2]{\;^{#1}\Box_{#2}^*\;}
\newcommand{\Baibj}{\B{\alpha i}{\beta j}}
\newcommand{\Bsaibj}{\Bs{\alpha i}{\beta j}}
\newcommand{\R}[2]{\,^{#1}\!\!\!\!\mbox{\footnotesize{R}}_{#2}}
\newcommand{\J}[2]{\,^{#1}\!\!\!\!\mbox{\footnotesize{J}}_{#2}}
\def\Pij{{\Phi_{ij}}}
\def\Pji{{\Phi_{ji}}}
\def\Pe{{\Phi_{23}}}
\def\Pu{{\Phi_{13}}}
\def\Pt{{\Phi_{12}}}
\newcommand {\V}[1] {V_{#1}}
\newcommand {\Vs}[1] {V^*_{#1}}
\def\Vai{{V_{\alpha i}}}
\def\Vajs{{V_{\alpha j}^*}}
\def\Vbj{{V_{\beta j}}}
\def\Vbis{{V_{\beta i}^*}}
\newcommand {\co}[1] {c_{#1}}
\newcommand {\s}[1] {s_{#1}}
\newcommand {\ct}[1] {c^2_{#1}}
\newcommand {\st} [1] {s^2_{#1}}
\def\ssp{\hspace{0.3 in}}
\def\vsp{\vspace{0.5 cm}}
\newcommand {\combin}[2]
  {\left( \stackrel{\raisebox{0.8ex}{$\displaystyle {#1}$}}
  {\raisebox{-1.3ex}{$\displaystyle {#2}$}} \right)}
%

\section{Introduction}

If neutrinos have mass, then the states of definite mass, $\nui$, may be 
distinct from the states
of definite flavor, $\nua$.  (We will use latin indices $i$, 
$j$, $k$, \ldots to refer to
mass states, and greek indices $\alpha$, $\beta$, $\gamma$, \ldots to refer 
to flavor states.)  
If the eigenvalues of the neutrino mass matrix are non-degenerate, then
neutrinos may change flavors, or oscillate, as they propagate.  The
long-standing solar neutrino deficit \cite{solar}, the atmospheric
neutrino anomaly \cite{atmos}, and the recent results from the LSND
experiment \cite{LSND} can all be understood in terms of oscillations between
neutrinos.  Traditional oscillation analyses generally assume
no more than three generations of neutrinos.  But resonant oscillations 
for the sun,
oscillations for the atmosphere, and the LSND data each require a different
neutrino mass-squared difference in neutrino oscillations are to 
account for all features of the data
\cite{BWW}.  Since three-neutrino models can have at most two independent
mass-squared differences, a sterile neutrino is apparently needed to reconcile 
all the data while retaining consistency with LEP measurements of
$Z \rightarrow \nu \overline{\nu}$ \cite{LEP}.  Several four-neutrino analyses
appear in the literature \cite{BWW,models}.  If, however, 
statistical or systematic errors
in the data evolve in the future, or if some data turns out to have an
explanation other than neutrino oscillations, then three-neutrino oscillations
may be sufficient.  One recent analysis of the complete set of data in the
three-neutrino framework is given in reference~\cite{TSI}.

Oscillation 
probabilities depend on products of four mixing-matrix elements.  Several
parameterizations of the mixing matrix in terms
of rotation angles have been introduced, beginning with the pioneering work of
Kobayashi and Maskawa \cite{KM}.  When three or
more neutrino generations are included in the oscillations analysis, the
oscillation  
probabilities become complicated functions of the 
neutrino mixing angles.  But oscillations are
observable and therefore parameterization-invariant.  One must ask if there is
not a better description of oscillations which avoids the arbitrariness of
angular-parameterization schemes.  In this paper, we introduce a new ``box'' 
parameterization of neutrino 
mixing valid for any number of neutrino generations.  
Oscillation probabilities are linear in the boxes, enabling a straighforward
analysis of oscillation data.  In what follows, we 
develop the algebra of the boxes and the unitarity constraints on that algebra. 
To conclude we illustrate the boxes' reduction to a basis in the case of three
generations.  A phenomenological analysis of existing oscillation data will be
performed in a future publication.


\section{The Standard Formulation of Neutrino Oscillations}
\label{3waysec}


The probability for a neutrino to oscillate from $\nua$ to $\nub$ is given by
the square of the transition amplitude:
\beq
\Pab =  \left|\sum_{i=1}^{n} \Vai {\Vbis}
e^{-i \phi_i} \right|^2 = 
\sum_{i=1}^{n} \sum_{j=1}^{n} (\Vai \;
\Vbis \; \Vajs \; \Vbj ) e^{-i \left( \half \Pij\right)}, 
\label{Probab}
\eeq
where $n$ is the number of neutrino generations and
\beq
\Phi_{ij} \equiv \half \left( \phi_i - \phi_j \right) = \half \left(
E_i t_i - p_i x_i - E_j t_j + p_j x_j \right).
\eeq
For relativistic neutrinos, $\Phi_{ij}$ is given by
\beq
\Phi_{ij} \approx \frac{\Delta m_{ij}^2}{4p} x, \mbox{\ \ where \ \ }
\Delta m_{ij}^2 \equiv m_i^2-m_j^2.
\label{Phiij}
\eeq
With a little bit of algebra, the oscillation probability 
may be brought into the form \cite{KimPev}
\beq
\Pab = \sum_{i=1}^n |\Vs{\alpha i}|^2 |\V{\beta i}|^2 + 2 \mbox{Re} \left[
  \sum_{i=1}^n \sum_{j \neq i} \Vs{\alpha j} \V{\beta j} 
  \V{\alpha i} \Vs{\beta i} e^{-i\Phi_{ij}}\right],
\label{Kimprob}
\eeq
or equivently, the form \cite{dissert}
\bea
\Pab & = & -2 \sum_i \sum_{j \neq i} \mbox{Re}(\Vai \Vbis \Vajs \;\Vbj) 
\sin^2 \left( \Phi_{ij} \right) 
\label{oscillation}  \\
&&+ \ \sum_i \sum_{j \neq i} \mbox{Im}(\Vai \Vbis \Vajs \;\Vbj)  
\sin \left( 2 \Phi_{ij} \right) + \delta_{\ab}.
\nonumber 
\eea
The probability for an antineutrino to oscillate from $\overline{\nu}_{\alpha}$ 
to $\overline{\nu}_{\beta}$ is obtained from $\Pab$ by replacing 
$V$ with $V^*$.  This is
equivalent to changing the sign of $\Pij$. 
$\Pabbar$ therefore differs from $\Pab$ merely in
the sign of the second term in equation~(\ref{oscillation}) or of the
exponent in equation~(\ref{Kimprob}).  Note that
$\Pab=\Pbabar$, as required by CPT-invariance.

Let us first make contact with the familiar case of two neutrino flavors, $n=2$.
 An
arbitrary $2\times2$ unitary matrix will have one rotation angle parameterizing
the real degree of freedom, and three phases.
But all three phases may be absorbed into the definitions of
Dirac fermion fields (or will cancel in oscillation probabilities for Majorana
neutrinos \cite{dissert,BP}), 
so a $2\times2$ mixing matrix V  has the simple form of a rotation matrix:
\beq
V = \left( 
\begin{array}{cc}
       \mbox{cos} \,\theta    &   -\mbox{sin} \,\theta  \\
       \mbox{sin} \,\theta    &    \mbox{cos} \,\theta
\end{array}
\right).
\label{twoflavorV}
\eeq
Since the matrix is explicitly real, the term in equation (\ref{oscillation})
involving imaginary matrix elements disappears, and the oscillation probability
in the two-flavor case is 
\beq
\Pab = \mbox{sin}^2 \,2 \theta \; \sin^2 \left( \frac{\Delta m^{2}_{12}}{4 p}
x \right) + \delta_{\ab}, \mbox{\ \ \ }n=2.
\label{twoflavorP}
\eeq
This result is  simple, and the mixing-angle parameterization is a natural
choice in the two-flavor situation.

The math becomes more complicated when we include three generations.
An arbitrary $3 \times 3$ unitary matrix has three real
degrees of freedom and six phases.  
$2n-1=5$  phases are not observable \cite{dissert,Nact}, 
since this is the number of
phase differences among the fermion fields which may be absorbed into field
redefinitions.
The three-generation mixing matrix  
is therefore described by three mixing angles and one phase.
The original choice of these four parameters, by Kobayashi and Maskawa to
describe quark mixing, 
is perhaps the best known parameterization.  Their choice of $V$ is
\cite{KM}
\beq
\left( \ba{ccc} 
\co{1} & \;\s{1} \co{3}\; & \s{1} \s{3} \\
-\s{1}\co{2} & \; \co{1}\co{2}\co{3}-\s{2}\s{3}e^{i\delta}\; &
\co{1}\co{2}\s{3}+\s{2}\co{3}e^{i\delta} \\
-\s{1}\s{2} & \co{1}\s{2}\co{3}+\co{2}\s{3}e^{i\delta} &
\co{1}\s{2}\s{3}-\co{2}\co{3}e^{i\delta}  \ea \right),
\label{NactCKM}
\eeq
where $\co{a}\equiv \cos \theta_a$, and $\s{a} \equiv \sin \theta_a$.  We will
refer to this particular choice of angles as the ``standard'' or ``KM''
parameterization.

In the standard KM parameterization, the phase only appears in the lower
right-hand sub-block of the matrix.  There is arbitrariness associated with the 
placement of the phase, since we absorb five relative phases into the field
definitions. 
Clearly the location of the phase cannot be
measurable.  Indeed, the CP-violating effects of the phase are contained in 
a single function of the phase called the Jarlskog invariant ${\cal J}$, 
which in the standard KM parameterization has the form \cite{Jarls1} 
${\cal J} = \co{1} \st{1} \co{2} \s{2} \co{3} \s{3} \sin {\delta}$.
Because of the arbitrariness of the phase convention, the phases of individual
matrix elements are not observable; only the magnitudes of matrix elements are
observable.

The observable oscillation probabilities 
are quite complicated functions of the mixing angles in angle-based 
parameterizations.  As an example, consider the product 
$\V{22}\Vs{23}\Vs{32}\V{33}$ appearing in the $\numu \rightarrow \nutau$ 
oscillation probability.  In the standard KM mixing matrix, this product 
is given by
\bea
\V{22}\Vs{23}\Vs{32}\V{33} &=& \ct{3} \st{3} \left[ \st{2} \ct{2} (s_{1}^4 + 6
\ct{1} + 2 \ct{1} \cos 2 \delta) - \ct{1} \right]
\nonumber \\
 & &  + \frac{{\cal J}}{\st{1}} (1+\ct{1})(\ct{2}-\st{2})(\st{3} - 
\ct{3}) \cot \delta + i {\cal J}, \mbox{\ \ \ }n=3.
\label{stanbox} 
\eea
Besides being ugly, thereby motivating the often-made 
two-generation approximation, the expression (but not its value) 
on the right-hand side of equation (\ref{stanbox}) is convention-dependent.  
Other parameterizations of the
unitary mixing matrix involving three angles and a phase are equally valid and
yield similarly complex oscillations probabilities.
Our development of a model-independent parameterization
has been motivated by the arbitrariness and complexity of the traditional
approach.


\section{A New Parameterization}
\label{boxes}


\subsection{Boxes Defined, and Their Symmetries}
\label{boxdefsec}

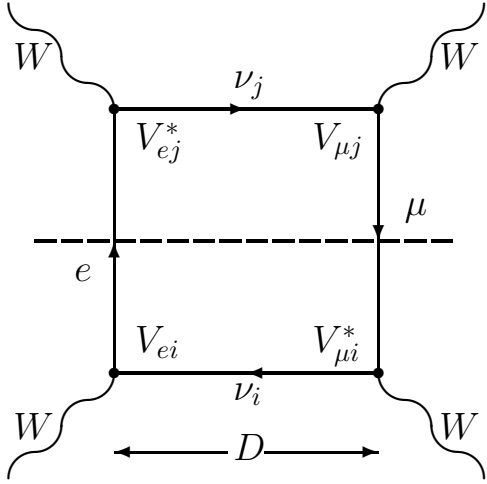
\begin{figure}[thb]
\begin{centering}
\begin{picture}(220,200)(0,-20)
\thicklines
\put(60,35){\vector(0,1){50}}
\put(60,85){\line(0,1){50}}
\put(60,135){\vector(1,0){50}}
\put(110,135){\line(1,0){50}}
\put(160,135){\vector(0,-1){50}}
\put(160,85){\line(0,-1){50}}
\put(160,35){\vector(-1,0){50}}
\put(110,35){\line(-1,0){50}}
\multiput(30,85)(10,0){16}{\line(1,0){8}}
\multiput(60,35)(100,0){2}{\circle*{4}}
\multiput(60,135)(100,0){2}{\circle*{4}}
\multiput(170,135)(20,20){2}{\oval(20,20)[tl]}
\multiput(170,155)(20,20){2}{\oval(20,20)[br]}
\multiput(170,35)(20,-20){2}{\oval(20,20)[bl]}
\multiput(170,15)(20,-20){2}{\oval(20,20)[tr]}
\multiput(50,35)(-20,-20){2}{\oval(20,20)[br]}
\multiput(50,15)(-20,-20){2}{\oval(20,20)[tl]}
\multiput(50,135)(-20,20){2}{\oval(20,20)[tr]}
\multiput(50,155)(-20,20){2}{\oval(20,20)[bl]}

\put(103,5){\vector(-1,0){43}}
\put(117,5){\vector(1,0){43}}

\Large

\put(45,70){$e$}
\put(105,143){$\nu_j$}
\put(170,95){$\mu$}
\put(105,25){$\nu_i$}
\put(105,2){$D$}
\put(68,43){$\V{ei}$}
\put(68,120){$\Vs{ej}$}
\put(135,43){$\Vs{\mu i}$}
\put(135,120){$\V{\mu j}$}
\put(183,150){$W$}
\put(183,10){$W$}
\put(22,10){$W$}
\put(22,150){$W$}

\normalsize

\end{picture}
\caption{Interference diagram contributing to the box
$^{ei}\Box_{\mu j}$, which appears in the 
$\nu_e \rightarrow \nu_{\mu}$ oscillation probability,
as the neutrino travels a distance $D$.  
Arrows on the diagram represent fermion number, not
momentum.}
\label{feynboxfig}
\end{centering}
\end{figure}

The immeasurability of the mixing matrix elements in the quark sector has been
addressed by numerous authors, such as those of references 
\cite{Jarls1,NP,Wu,BD,DDW}.  Measurable quantities include only the
magnitudes of mixing matrix elements, the products of four mixing-matrix
elements appearing in the oscillation probabilities, and
particular higher-order functions of mixing-matrix elements \cite{Wu,KS}.  
As evidenced in equations~(\ref{Probab}), (\ref{Kimprob}), and
(\ref{oscillation}), neutrino oscillation probabilities depend linearly on 
the fourth-order objects,
\beq
\Baibj \equiv \Vai V^{\dagger}_{i \beta} \Vbj V^{\dagger}_{j \alpha} 
= \Vai \V{\alpha j}^* \V{\beta i}^* \Vbj,
\label{boxdef}
\eeq
which we call ``boxes'' since each contains as factors the 
corners of a submatrix, or ``box," of the mixing matrix.
For example, the upper left $2\times 2$ submatrix elements produce the box
\beq
\B{11}{22} = \V{11} \V{12}^* \V{21}^* \V{22}.
\eeq
These boxes are the neutrino equivalent of the ``plaques'' used by 
Bjorken and Dunietz for another purpose in the quark sector in reference
\cite{BD}.  The name ``box'' also seems appropriate in light of the Feynman
diagram which describes the oscillation process (Figure~\ref{feynboxfig}).

In general, each of the box indices $i$, $j$, $\alpha$, and $\beta$ may be any
number between $1$ and $n$, the number of neutrino flavors.  We therefore
initially have $n^4$ possible boxes.
Examination of equation~(\ref{boxdef}), however, 
reveals a few symmetries in the indexing:
\beq
\Baibj = \B{\beta j}{\alpha i} = \Bs{\beta i}{\alpha j} = 
\Bs{\alpha j}{\beta i}.
\label{symmetries}
\eeq
If the order of either set of indices is reversed ({\it id est}, 
$j \leftrightarrow i$
or $\beta \leftrightarrow \alpha$), the box turns into its complex conjugate; 
if both sets of indices are reversed, the box returns to its original value
\cite{NP}. 
And if $V$ is replaced by $V^{\dagger}$, then
$\Baibj\rightarrow \Bs{i\alpha}{j\beta}$.
 

\subsection{Degenerate and Nondegenerate Boxes}

 
Equation~(\ref{symmetries}) demonstrates that boxes with $\alpha = \beta$ or 
$i = j$, are real.  Indeed, these are given from equation~(\ref{boxdef}) as
\beq
\B{\alpha i}{\alpha j} = |\V{\alpha i}|^{2} |\V{\alpha j}|^{2}, 
\mbox{\ \ and \ \ } 
\B{\alpha i}{\beta i} = |\V{\alpha i}|^{2} |\V{\beta i}|^{2}.
\label{samei} 
\eeq
Those boxes with both sets of indices equal are
\beq
\B{\alpha i}{\alpha i} = |\Vai|^4.
\eeq
We call boxes with one and two repeated indices ``singly-degenerate'' and
``doubly-degenerate,'' respectively.  Boxes with $\alpha \neq \beta$ and $i\neq
j$ are called nondegenerate.
As can be seen from equation (\ref{oscillation}), singly-degenerate boxes 
with repeated flavor indices enter into the formulae for flavor-conserving
survival probabilities, but not for flavor-changing transition
probabilities.  Degenerate boxes with repeated mass indices (including 
the doubly-degenerate boxes) do not appear in
any oscillation formula.  Degenerate boxes may be expressed in terms of 
the nondegenerate boxes, as will be shown shortly. This
possibility and the symmetries expressed in 
equation~(\ref{symmetries}) allow us to express combinations of boxes in terms
of only the    
nondegenerate ``ordered'' boxes for which $\alpha < \beta$ and    
$i < j$. 

The number of flavor-index pairs satisfying $\alpha < \beta$ ordering is 
$N \equiv \combin{n}{2}= 
\frac{n\left(n-1\right)}{2}$.  $N$ 
mass-index
pairs similarly satisfy $i<j$ ordering.  Thus, the number of ordered
nondegenerate boxes is $N^2$.  The number of doubly-degenerate boxes is $n^2$.
The number of ordered singly-degenerate boxes
is $\half n^2 \left(n-1\right) = nN$ for flavor-degenerate boxes,
and the same for mass-degenerate boxes, yielding $2nN$ total singly-degenerate
ordered boxes.  For $n=3$, $N=3$ too, and
we have nine ordered nondegenerate boxes, nine ordered singly-degenerate boxes,
and nine doubly-degenerate boxes.  
For $n>3$, matrix elements $\V{\alpha i}$ with $\alpha >3$ are not easily
accessible since $\alpha >3$ labels heavy charged leptons beyond the usual $e$,
$\mu$, $\tau$.  Therefore, in the absence of processes with sufficient energy to
excite the heavy charged lepton degree of freedom, boxes with flavor indices
exceeding $3$ are not physically accessible either.  The number of accessible
ordered nondegenerate boxes is $3N$.  The number of accessible ordered
singly-degenerate boxes is $3N$ with flavor degeneracy and $3n$ with mass
degeneracy, yielding $3(n+N)$ for the total.
These counts are recapped
later in Table~\ref{counttab}, along with other box counts.


\subsection{Oscillation Probabilities in Terms of Boxes}

 
Using the symmetries expressed in equation~(\ref{symmetries}), equation 
(\ref{oscillation}) becomes
\beq
\Pab  =  -2 \sum_{i=1}^{n} \sum_{j>i} \left[ 2 \R{\alpha i}{\beta j}
\sin^2 \Pij - \J{\alpha i}{\beta j} \sin 2 \Pij \right] +
\delta_{\alpha \beta},
\label{boxorig}
\eeq
where we have defined the shorthand $\R{\alpha i}{\beta j} \equiv
\mbox{Re}\left(\Baibj\right)$ and $\J{\alpha i}{\beta j}
\equiv\mbox{Im}\left(\Baibj\right)$.
When individual oscillations cannot be resolved due to uncertainties in energy
or position, the $\sin^2\Pij$ and $\sin 2\Pij$ are effectively averaged, giving
the transition probabilities
\beq
\langle \Pab \rangle = -2\sum_{i=1}^{n} \sum_{j>i}\R{\alpha i}{\beta j}.
\eeq
These probabilities are equivalently given by
\beq
\langle \Pab \rangle
= \sum_{i=1}^n |\Vai|^2 |\V{\beta i}|^2 = \sum_{i=1}^n \B{\alpha i}{\beta i},
\eeq
which may be a more familiar expression; the equivalence of these expressions
will be shown later in this paper.

The survival probabilities $\Paa$ may be found from the transition
probabilities $\Pab$ by
\beq
\Paa = 1 - \sum_{\beta \neq \alpha} \Pab.
\eeq
Survival probabilities are more simply expressed in terms of degenerate boxes,
or $|V|$s, rather than nondegenerate boxes.  They are
\beq
\Paa =  1-4\sum_{i=1}^n \sum_{j>i} \B{\alpha i}{\alpha j} \sin^2\Pij =
1-4\sum_{i=1}^n \sum_{j>i} |\V{\alpha i}|^2 |\V{\alpha j}|^2 \sin^2\Pij.
\eeq
If oscillations are averaged, the survival probability is
\beq
\langle\Paa \rangle =1-2\sum_{i=1}^{n}\sum_{j>i}|\V{\alpha i}|^2 |\V{\alpha
j}|^2
= 1-\sum_{i=1}^{n}\sum_{j\neq i}|\V{\alpha i}|^2 |\V{\alpha j}|^2
= \sum_{i=1}^{n} |\V{\alpha i}|^4 = \sum_{i=1}^{n} \B{\alpha i}{\alpha i}.
\eeq
The third equality here results from unitarity of $V$.

Interchanging $\alpha \leftrightarrow \beta$ in equation (\ref{boxorig}) gives
the
time-reversed reactions $\Pba$:
\beq
\Pba =  -2 \sum_{i=1}^{n} \sum_{j>i} \left[ 2 \R{\alpha i}{\beta j}
\sin^2 \Pij + \J{\alpha i}{\beta j} \sin 2 \Pij \right] +
\delta_{\alpha \beta},
\eeq
so a measure of T-violation, or equivalently CP-violation, in the neutrino
sector is 
\beq
\Pab - \Pba = \Pab - \Pabbar = 
4 \sum_{i=1}^{n} \sum_{j>i}\J{\alpha i}{\beta j} \sin 2 \Pij.
\label{CPprob}
\eeq

Ignoring possible CP-violating phases in the mixing matrix, 
there are $\combin{n}{2}=N$ real parameters determining $V$.  
Determining these $N$ parameters determines the complete mixing matrix.  
Conveniently, there are $N$ transition probabilities
$\Pab=\Pba$.
Thus, all of the information in the mixing matrix is contained in the $N$
transition probabilities.  In this sense, they form a convenient basis for
determining all oscillation parameters.  Of course, if the same transition
probability is measured at two or more different distances, then all $N$
transition probabilities may not be needed to determine $V$.

Allowing CP-violation in the mixing matrix, there are $N$ real parameters and
$\half (n-1)(n-2)$ phases, for a total of $(n-1)^2$ parameters.  With
CP-violation, however, there are $2N=n(n-1)$ independent transition
probabilities $\Pab$.  The number of transition probabilities exceeds the
number of independent parameters, so they again form a convenient
basis for determining the mixing matrix.  
In reality, only the three flavor indices $e$, $\mu$, $\tau$ are easily
accessible, as alluded to at the end of the previous section.  Moreover, some
of the $N$ parameters in the mixing matrix, namely those which rotate sterile
states for $n\ge 5$, are not accessible.  The inaccessibility issue complicates
the counting.  We do not pursue it further here.
 
The transition probabilities for which $\alpha \neq \beta$ in 
equation (\ref{boxorig}) may be expressed in matrix form.  The matrix of boxes 
will
necessarily be an $N\times N$ matrix since the vector on the left-hand side and
the sum on the right-hand side of equation (\ref{boxorig}) each contain $N$
components or terms.  For three flavors, 
we have
\beq
{\cal P}(n=3) \equiv
\left( \ba{c}
\Peu \\ \Put \\ \Pet
\ea \right) =  
-4 \mbox{ Re}({\cal B})\; S^2(\Phi) + 2 \mbox{ Im} ({\cal B})\; S(2\Phi),
\label{boxprob}
\eeq
where
\beq
{\cal B} \equiv 
\left( \ba{ccc}
\B{e1}{\mu 2} & \B{e2}{\mu 3} & \B{e1}{\mu 3} \\
\B{\mu 1}{\tau 2} & \B{\mu 2}{\tau 3} & \B{\mu 1}{\tau 3} \\
\B{e1}{\tau 2} & \B{e2}{\tau 3} & \B{e1}{\tau 3}
\ea \right), \mbox{\ \ and \ \ } 
S^k(\Phi) \equiv  
\left( \ba{c}
\sin ^k\Pt \\  \sin^k \Pe \\ \sin^k \Pu
\ea \right), \mbox{\ \ \ }n=3.
\label{boxbox}
\eeq
For the time-reversed channels, or for the antineutrino channels, the sign of
the Im$({\cal B})$ term is reversed.
The definitions of $\cal{P}$, $\cal{B}$ and $S^k(\Phi)$ may be extended to any 
number of flavors.  Adding a fourth, perhaps
sterile, neutrino flavor increases the dimensions of ${\cal B}$ to $6 \times 6$,
since there are now six independent transition probabilities and
six mass differences.  
The number of ordered nondegenerate boxes increases
from nine to thirty-six, but only eighteen of them are accessible.  
For the six flavors of neutrinos in mirror-symmetric
schemes, ${\cal B}$ is a
fifteen-by-fifteen matrix, and we have two hundred twenty-five ordered
nondegenerate boxes.  But the number of accessible boxes is just forty-five.

Our parameterization is
especially well-suited for considering higher numbers of generations.  
While using boxes in $n>3$ situations may be cumbersome, it is much less so than
extending a mixing-angle parameterization to higher generations.  
Our matrix ${\cal B}$ of boxes merely acquires extra
columns when new flavors are introduced; extra rows are not accessible at
energies below new charged-lepton thresholds.  Furthermore, 
oscillation probabilities are linear in boxes, no matter how many generations.


\subsection{A Return to Two Generations}


The boxes may be used to illustrate when the two-flavor oscillation
formula~(\ref{twoflavorP}) is a meaningful approximation and when it is not. 
For this, we assume ordering of the mass indices $m_1<m_2<\cdots <m_n$.  If the
transit distance $L$ of the neutrino is just large enough to impact upon the
shortest oscillations length $\frac{4\pi p}{\Delta m_{n1}^2}$,
but not upon the next-shortest, then 
\beq
\sin^2 2\theta_{\alpha \beta}^{eff} = -4\R{\alpha 1}{\beta n}
\eeq 
is
meaningful.  If one amplitude $\R{\alpha i}{\beta j}$ among those associated
with $\Delta m_{ij}^2$ satisfying $\Delta m_{ij}^2 \gsim \frac{4\pi p}{L}$ 
dominates the others, then 
\beq
\sin^2 2\theta_{\alpha \beta}^{eff} = -4\R{\alpha i}{\beta j}
\eeq
is a meaningful approximation.  If there is a dominant mass 
$m_n \gg m_{n-1}, \cdots, m_1$, then for 
$\frac{4\pi p}{m_n^2} \lsim L \ll \frac{4\pi p}{\Delta m_{n-1, 1}^2}$, 
\beq
\sin^2 2\theta_{\alpha \beta}^{eff} = -4\sum_{i=1}^{n-1}
  \R{\alpha i}{\beta n}
\label{2genapprox}
\eeq
is meaningful.  We will show later (equation~(\ref{1degcol})) that 
the right-hand side of equation~(\ref{2genapprox}) may be written as 
$4\left|\V{\alpha n}\right|^2\left|\V{\beta n}\right|^2$.
If the largest masses are nearly
degenerate, with degeneracy $q$, then the above becomes
\beq
\sin^2 2\theta_{\alpha \beta}^{eff} = -4\sum_{i=1}^{n-q} \ \ \sum_{j=n-q+1}^n
  \R{\alpha i}{\beta j},
\eeq
again meaningful.  However, if two or more well-separated $\Delta m_{ij}^2$s
exceed $\frac{4 \pi p}{L}$, then the two-flavor formula is inapplicable.


\subsection{Relations between Mixing Matrix Elements and Boxes}
\label{boxVsec}


Neutrino oscillation experiments will directly measure the boxes in equation 
(\ref{boxorig}),  
not the individual mixing matrix elements, $\Vai$.  But one would like to obtain
the fundamental $\Vai$ from the measured boxes.  We develop here an algebra
relating boxes and mixing matrix elements.

Some tautologous relationships
between the degenerate and nondegenerate boxes are easily confirmed using
equation  (\ref{boxdef}); they hold for any number of generations:
\bea
|V_{\alpha i}|^2 |V_{\alpha j}|^2 & = & \B{\alpha i}{\alpha j} =
   \frac{\Bs{\alpha i}{\eta j} \B{\alpha i}{\lambda j}}{\B{\eta i}{\lambda j}},
 \;\;\;\;\;\;\;\;\;\;\;\;
 (\eta \neq \lambda \neq \alpha), 
 \label{vproda} \\
\nonumber \\
|V_{\alpha i}|^2 |V_{\beta i}|^2 & = & \B{\alpha i}{\beta i} = 
  \frac{\Bs{\alpha i}{\beta x} \B{\alpha i}{\beta y}}{\B{\alpha x}{\beta y}}, 
 \;\;\;\;\;\;\;\;\;\;\;
 (x \neq y \neq i),  \mbox{\ \ and \ \ } 
 \label{vprodi} \\
\nonumber \\
\frac{|\V{\alpha i}|^2}{|\Vbj|^2} & = & 
   \frac{\Bs{\alpha i}{\eta j}\B{\alpha i}{\beta x}}
   {\B{\alpha j}{\beta x} \Bs{\beta i}{\eta j}}, 
 \;\;\;\;\;\;\;\;\;\;(\eta \neq \alpha \neq \beta,
 \mbox{ and } x \neq i \neq j).
 \label{vratio}
\eea
In these equations and what follows, $\alpha$, $\beta$, $\gamma$, $i$, $j$, and
$k$ will usually 
be reserved for indices that are chosen at the start of a calculation, 
while other indices such as $x$, $y$, $\eta$, and $\lambda$ primarily 
represent ``dummy''
indices which are chosen arbitrarily except to respect the inequality 
constraints following equations such as equation (\ref{vratio}).  

\begin{figure}[htb]
\vsp
\begin{centering}
\begin{picture}(230,100)(0,-30)
\thicklines
\multiput(5,65)(30,0){3}{\circle*{4}}
\multiput(5,35)(30,0){3}{\circle*{4}}
\multiput(5,5)(30,0){3}{\circle*{4}}
\put(5,61){\vector(0,-1){22}}
\put(35,39){\vector(0,1){22}}
\multiput(165,65)(30,0){3}{\circle*{4}}
\multiput(165,35)(30,0){3}{\circle*{4}}
\multiput(165,5)(30,0){3}{\circle*{4}}
\put(169,65){\vector(1,0){22}}
\put(191,35){\vector(-1,0){22}}
\put(30,-25){(a)}
\put(190,-25){(b)}
\end{picture}
\vsp
\caption{The graphical representation for 
(a) the box $\B{11}{22}$, and (b) its inverse $\left(\B{11}{22}\right)^{-1}$.  
Vertical arrows point from the
matrix elements which are not complex conjugated in the box to the 
complex-conjugated elements.  While shown here in three generations, this
representation may be extended to higher generations merely by adding more dots.
\label{b1122fig}}
\end{centering}
\end{figure}
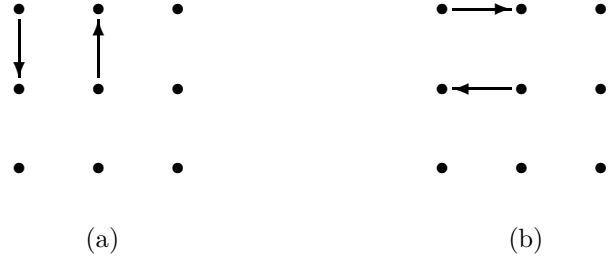

The tautologies (\ref{vproda}) to (\ref{vratio}) become evident not only by
algebraically using the definitions of the boxes, but also by considering
a graphical representation we have developed.  This representation is
discussed in detail in Appendix~\ref{graphapp}, and it 
proves quite useful when considering relationships 
between boxes.  Figure~\ref{b1122fig} illustrates the representations of 
$\B{11}{22}$  and $\left(\B{11}{22}\right)^{-1}$.  

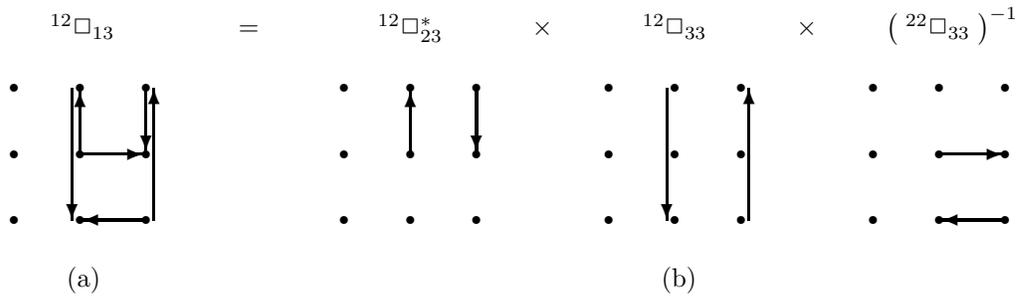
\begin{figure}[thb]
\vsp
\vsp
\begin{picture}(425,105)(0,95)
\thicklines
\multiput(25,175)(25,0){3}{\circle*{3}}
\multiput(25,150)(25,0){3}{\circle*{3}}
\multiput(25,125)(25,0){3}{\circle*{3}}
\put(47,175){\vector(0,-1){50}}
\put(50,151){\vector(0,1){23}}
\put(78,125){\vector(0,1){50}}
\put(75,174){\vector(0,-1){23}}
\put(51,150){\vector(1,0){23}}
\put(74,125){\vector(-1,0){23}}
\put(36,195){$\B{12}{13}$}
\put(45,100){(a)}
\multiput(150,175)(25,0){3}{\circle*{3}}
\multiput(150,150)(25,0){3}{\circle*{3}}
\multiput(150,125)(25,0){3}{\circle*{3}}
\put(175,151){\vector(0,1){23}}
\put(200,174){\vector(0,-1){23}}
\multiput(250,175)(25,0){3}{\circle*{3}}
\multiput(250,150)(25,0){3}{\circle*{3}}
\multiput(250,125)(25,0){3}{\circle*{3}}
\put(272,175){\vector(0,-1){50}}
\put(303,125){\vector(0,1){50}}
\multiput(350,175)(25,0){3}{\circle*{3}}
\multiput(350,150)(25,0){3}{\circle*{3}}
\multiput(350,125)(25,0){3}{\circle*{3}}
\put(376,150){\vector(1,0){23}}
\put(399,125){\vector(-1,0){23}}
\put(110,195){$=$}
\put(160,195){$\Bs{12}{23}$}
\put(221,195){$\times$}
\put(260,195){$\B{12}{33}$}
\put(321,195){$\times$}
\put(355,195){$\left(\B{22}{33}\right)^{-1}$}
\put(270,100){(b)}
\end{picture}
\vsp
\caption{The graphical representation in three generations for 
(a) the degenerate box $\B{12}{13}$, \ and 
(b) the combination of nondegenerate boxes which ``create'' it.  
\label{vprodafig}}
\end{figure}

To illustrate an example of equation~(\ref{vproda}), the degenerate 
box $\B{12}{13}$ is ``created'' in Figure~\ref{vprodafig}a 
when one uncanceled vertical
arrow enters and another leaves each of the matrix elements
$V_{12}$ and $V_{13}$, while arrows at all other points cancel.  Those
arrows are the result of the combination of ordered boxes given in
(\ref{vproda}), as shown in Figure~\ref{vprodafig}b.


Equations (\ref{vproda}) and (\ref{vprodi}) 
are themselves special cases of the more general
\bea
\label{2boxij}
\Baibj \B{\gamma i}{\delta j} &=& \left[ \Vai \Vajs \Vbj \Vbis \right]
\left[ V_{\gamma i} \Vs{\gamma j} V_{\delta j} \Vs{\delta i} \right] \\
& = &\left[ \Vai \Vajs V_{\delta j} \Vs{\delta i}\right]
\left[ V_{\gamma i} \Vs{\gamma j} \Vbj \Vbis \right] = 
\B{\alpha i}{\delta j} \B{\gamma i}{\beta j}, \nonumber
\eea
and the analogous relation
\beq
\Baibj \B{\alpha k}{\beta l} = \B{\alpha i}{\beta l} \B{\alpha k}{\beta j}.
\label{2boxab}
\eeq
These relations hold for both degenerate boxes and nondegenerate boxes. 
Including disordered boxes adds no new information, so we will as usual consider
only ordered boxes.  

This index rearrangement of equations~(\ref{2boxij}) and (\ref{2boxab}) may
straightforwardly be extended to obtain higher-order relationships (trilinear,
quadrilinear, {\it et cetera}) in the boxes.  These higher-order relationships
do not enter into oscillation probabilities, so we will not
examine them here.

We may express $|V_{\alpha i}| = \left(\B{\alpha i}{\alpha
i}\right)^{\frac{1}{4}}$ 
in terms of three singly-degenerate boxes by setting $\alpha=\beta$ 
in equation~(\ref{vprodi}).  Then, using equation (\ref{vproda})
to substitute for the singly-degenerate boxes yields an expression for the
doubly-degenerate box in terms of nine nondegenerate boxes: 
\bea 
|V_{\alpha i}|^4 = \B{\alpha i}{\alpha i} =  
\frac{\B{\alpha i}{\alpha x} \B{\alpha
i}{\alpha y}}{\B{\alpha x}{\alpha y}} = 
\frac{\B{\alpha x}{\tau i} \B{\alpha i}{\sigma x}  \B{\alpha y}{\rho i}
\B{\alpha i}{\zeta y} \B{\omega x}{\mu y}} {\B{\tau i}{\sigma x} \B{\rho
i}{\zeta y} \B{\alpha y}{\omega x}  \B{\alpha x}{\mu y}}, \;\;\;& 
\left\{
\ba{c}
\tau \neq \sigma \neq \alpha \\
\zeta \neq \rho \neq \alpha \\
\mu \neq \omega \neq \alpha \\
x \neq y \neq i 
\ea .
\right.
\label{vfour}
\eea
As noted earlier, flavor-degenerate boxes give the probabilities for
flavor-conserving oscillations measured in disappearance experiments, while 
nondegenerate boxes give the probabilities for flavor-changing oscillations
measured in appearance experiments.  Since the degenerate and nondegenerate
boxes are related, individual 
$|V_{\alpha i}|$ may be deduced from an appropriate 
set of measurements of either kind.

One can obtain a relationship similar to equation~(\ref{vfour}) 
by setting $i=j$ in
equation~(\ref{vproda}) and then using equation (\ref{vprodi}) to substitute
for the singly-degenerate boxes: \bea
|V_{\alpha i}|^4 = \B{\alpha i}{\alpha i} =
\frac{\B{\alpha i}{\lambda i} \B{\alpha i}{\eta i}}{\B{\lambda i}{\eta i}} = 
\frac{\B{\alpha i}{\lambda n} \B{\lambda i}{\alpha p} 
\B{\alpha i}{\eta r} \B{\eta i}{\alpha s} \B{\eta t}{\lambda u}}
{\B{\lambda n}{\alpha p} \B{\eta r}{\alpha s} \B{\lambda i}{\eta t} 
\B{\eta i}{\lambda u}}, \;\;\;& 
\left\{
\ba{c}
n \neq p \neq i \\
r \neq s \neq i \\
t \neq u \neq i \\ 
\lambda \neq \eta \neq \alpha 
\ea .
\right.
\label{vfour2}
\eea

\begin{figure}[htb]
\vsp
\begin{centering}
\begin{picture}(250,115)(-10,10)
\thicklines
\multiput(0,40)(40,0){3}{\circle*{4}}
\multiput(0,80)(40,0){3}{\circle*{4}}
\multiput(0,120)(40,0){3}{\circle*{4}}
\put(-10,118){\vector(0,-1){76}}
\put(-6,42){\vector(0,1){76}}
\put(-2,82){\vector(0,1){36}}
\put(2,118){\vector(0,-1){36}}
\put(38,82){\vector(0,1){36}}
\put(42,118){\vector(0,-1){76}}
\put(82,118){\vector(0,-1){36}}
\put(86,42){\vector(0,1){76}}
\put(38,78){\vector(0,-1){36}}
\put(78,42){\vector(0,1){36}}
\put(78,122){\vector(-1,0){36}}
\put(42,118){\vector(1,0){36}}
\put(2,82){\vector(1,0){76}}
\put(38,78){\vector(-1,0){36}}
\put(42,78){\vector(1,0){36}}
\put(2,42){\vector(1,0){36}}
\put(78,42){\vector(-1,0){36}}
\put(78,38){\vector(-1,0){76}}
\multiput(160,40)(40,0){3}{\circle*{4}}
\multiput(160,80)(40,0){3}{\circle*{4}}
\multiput(160,120)(40,0){3}{\circle*{4}}
\put(158,118){\vector(0,-1){36}}
\put(162,118){\vector(0,-1){36}}
\put(158,78){\vector(0,-1){36}}
\put(162,78){\vector(0,-1){36}}
\put(198,118){\vector(0,-1){76}}
\put(202,82){\vector(0,1){36}}
\put(202,42){\vector(0,1){36}}
\put(238,82){\vector(0,1){36}}
\put(238,42){\vector(0,1){36}}
\put(242,42){\vector(0,1){76}}
\put(162,122){\vector(1,0){76}}
\put(162,118){\vector(1,0){36}}
\put(202,118){\vector(1,0){36}}
\put(238,82){\vector(-1,0){36}}
\put(202,78){\vector(1,0){36}}
\put(198,42){\vector(-1,0){36}}
\put(238,42){\vector(-1,0){36}}
\put(238,38){\vector(-1,0){76}}
%
%
\put(35,15){(a)}
\put(195,15){(b)}
\end{picture}
\vsp
\caption[The graphical representation for 
(a) $|\V{11}|^4$, and \ (b) $|\V{21}|^4$.]{The graphical
representation for 
(a) $|\V{11}|^4$, and \ (b) $|\V{21}|^4$.  Notice that each matrix
point except the one being represented has an equal number of vertical arrows as
horizontal arrows entering and leaving.  The element being represented has two
vertical arrows leaving (representing $V_{\alpha i}^2$) and two entering
(representing $V_{\alpha i}^{*2}$) to produce $|\Vai|^2$.  
Appendix~\ref{graphapp} provides more details.
\label{V4fig}}
\end{centering}
\end{figure}
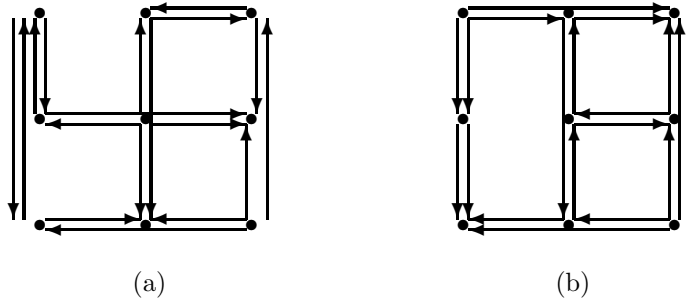

In the three-generation case, equations (\ref{vfour}) and (\ref{vfour2}) 
are uniquely specified by the index constraints in brackets, and they are 
equivalent to each
other since all nine boxes are used in both equations.
For example,
\beq
|\V{11}|^4 = 
\frac{\B{11}{22} \Bs{11}{23} \B{11}{33} \Bs{11}{32} \B{22}{33}}
{\Bs{12}{23} \B{12}{33} \B{21}{33} \Bs{21}{32}}
\label{V11four}
\eeq
with $\alpha=i=1$ holds with any number of generations, but it is the unique $5$
on $4$ box representation of $\left|\V{11}\right|^4$ in three generations.  
Another example in three generations is
\beq
|\V{21}|^4 = \frac{\B{21}{32} \B{11}{22} \B{21}{33} \B{11}{23} \B{12}{33}}
{\B{11}{32} \B{11}{33} \B{22}{33} \B{12}{23}},
\label{V21four}
\eeq 
again true for all $n$ and unique in $n=3$.
The equalities in equations~(\ref{V11four}) and (\ref{V21four}) are
illustrated in Figure~\ref{V4fig}.

If $n>3$, equations (\ref{vfour}) and (\ref{vfour2}) still hold, but the indices
of each are not specified uniquely.  
In higher generations the indices may take on values larger than 3; if $n=6$,
equation (\ref{vfour}) may contain up to six different flavor indices 
but only three mass indices, while equation (\ref{vfour2}) may contain
more mass indices and only three flavor indices.  

Before continuing, we should note that all of the relationships in this section 
follow from the definitions of the
boxes in equation (\ref{boxdef}) and so are valid for any matrix, unitary or
otherwise.  The constraints of unitarity, to which we will turn our attention
after we count the number of independent boxes, 
will provide us with expressions for $|\Vai|^4$ (found in
equations~(\ref{2degxy}) and (\ref{2degetlam}))
which are easier to manage than the expressions (\ref{vfour}) and (\ref{vfour2})
above.


\subsection{The Number of Independent Boxes}
\label{dogsec}


An $n \times n$ arbitrary matrix has $n^2$ elements, which may be complex,
leading to $2 n^2$ parameters.  The neutrino mixing matrix, however, 
is unitary, and has
$n^2$ unitarity conditions as constraints.  Of the $n^2$ remaining parameters,
$\half n(n-1)$ (the number of parameters in a real unitary, or orthogonal,
matrix) may be taken to be real rotation angles, 
and the remaining $\half n(n+1)$ are phases making the matrix complex. 
We may redefine the $2n$ neutrino and charged lepton fields to
remove $2n-1$ relative phases from 
the mixing matrix, leaving 
$\half n(n+1) - (2n-1) = \half (n-1)(n-2)$ phases measurable in oscillations.
The number of independent real and imaginary
box parameters should be the same as the number of real and
imaginary CKM parameters \cite{KS}.  As we will see, unitarity relates the
imaginary and real box parameters, so one may choose more real boxes and fewer
imaginary boxes as the independent basis of $(n-1)^2$ elements.  In fact, one
may choose a basis of only real boxes.  For $n>3$, not all of the boxes are
generally accessible (and not all of the $(n-1)^2$ mixing matrix parameters 
are accessible), so the basis of boxes may be smaller.  
The parameter counts for the mixing matrix and the boxes are
summarized in Table~\ref{counttab}.

\vfill\eject

\begin{table}[htb]
\caption{Parameter counting for the mixing matrix and boxes. 
$N$ is shorthand for the frequently-appearing combination 
$\protect \combin{n}{2} \equiv \half n(n-1)$.  
$n_f$ represents the number of active neutrino 
flavors, not to exceed $3$, and 
$N_f \equiv \protect \combin{n_f}{2}$. 
\label{counttab}}
\centering
\begin{tabular}{|c||c|c|c|c|c|} \hline
Number of generations        & $n$                & 2  & 3  & 4   & 6 
  \\ \hline \hline
Params. for arbitrary matrix & $2 n^2$            & 8  & 18 & 32  & 72 
  \\ \hline
Unitarity constraints        & $n^2$              & 4  & 9  & 16  & 36 
  \\ \hline
Relative phases              & $2n-1 $            & 3  & 5  & 7   & 11 
  \\ \hline
Real params for unitary V    & $N$                & 1  & 3  & 6   & 15 
  \\ \hline
Remaining phases in V        & $\half(n-1)(n-2)$  & 0  & 1  & 3   & 10
  \\ \hline
Physical params in V         & $(n-1)^2$          & 1  & 4  & 9   & 25
  \\ \hline \hline
Initial boxes                & $n^4$              & 16 & 81 & 256 & 1296
  \\ \hline
Doubly-degenerate boxes      & $n^2$              & 4  &  9 &  16 & 36
  \\ \hline
Ordered singly-degenerate boxes & $2nN$           & 4  & 18 &  48 & 180
  \\ \hline
Ordered nondegenerate boxes  & $N^2$              & 1  & 9  & 36  & 225
  \\ \hline
Independent Re$(\Baibj)$s    & $N$                & 1  & 3  & 6   & 15
  \\ \hline
Independent Im$(\Baibj)$s    & $\half(n-1)(n-2)$  & 0  & 1  & 3   & 10
  \\ \hline
Accessible boxes             & $n_f^2 n^2$        & 16 & 81 & 144 & 324
  \\ \hline
Accessible doubly-deg. boxes & $n_f n$            & 4  & 9  & 12  & 18
  \\ \hline
Accessible ordered singly-deg. boxes & $n_fN+N_fn$& 4  & 18 & 30  & 63
  \\ \hline
Accessible ordered nondeg. boxes  & $N_f N$       & 1  &  9 & 18  & 45
  \\ \hline
\end{tabular}
\end{table}


\section{Unitarity Relations Among the Boxes}


Unitarity requires  that
\bea
\sum_{\eta=1}^n \V{\eta i} \V{\eta j}^* & = & \delta_{ij}, 
\label{unitrow} \mbox{ \ \ and \ \ } \\
\sum_{y=1}^n \V{\alpha y} \V{\beta y}^* & = & \delta_{\alpha \beta}.
\label{unitcol}
\eea
These equations are not independent sets of constraints at this 
point:
equation~(\ref{unitrow}) states \mbox{$V^{\dagger}V=\openone$,} and
equation~(\ref{unitcol})
states \mbox{$VV^{\dagger}=\openone$.} 
We can, however, use these equivalent equations to obtain two separate sets of
constraints on boxes by multiplying equation (\ref{unitrow}) by 
$V_{\lambda i}^* V_{\lambda j}$ and equation (\ref{unitcol}) by 
$V_{\alpha x}^* V_{\beta x}$:
\beq
\hspace{1.0 cm} \sum_{\eta=1}^n \B{\eta i}{\lambda j} 
  =\sqrt{\B{\lambda i}{\lambda i}} \delta_{ij}, \mbox{\ \ and\ \ }
\label{urow} 
\eeq
\beq
\sum_{y =1}^n \B{\alpha y}{\beta x} 
  = \sqrt{\B{\alpha x}{\alpha x}}\delta_{\ab}.
\label{ucol}
\eeq
Isolating the manifestly degenerate boxes from the nondegenerate boxes,
equation~(\ref{urow}) becomes
\beq
\sum_{\eta \neq \lambda} \B{\eta i}{\lambda j} = 
\sqrt{\B{\lambda i}{\lambda i}} \delta_{ij} - \B{\lambda i}{\lambda j},
\label{prof1}
\eeq
and equation~(\ref{ucol}) becomes
\beq
\sum_{y\neq x} \B{\alpha y}{\beta x} = 
\sqrt{\B{\alpha x}{\alpha x}}\delta_{\ab} - \B{\alpha x}{\beta x}.
\label{prof2}
\eeq
One can show that equations~(\ref{prof1}) and (\ref{prof2}), although distinct,
are related by the symmetry that takes $\Baibj \rightarrow
\Bs{i\alpha}{j\beta}$, which we identified earlier as the symmetry $V\rightarrow
V^{\dagger}$.  We will find several pairs of equations related by this symmetry
which differ only in whether the
sum is over a flavor index or a mass index.

Summing equation~(\ref{urow}) over $\lambda$ in the $i\neq j$ case, we find
\beq
0=\sum_{\lambda=1}^n \sum_{\eta=1}^n \B{\eta i}{\lambda j} = 
\sum_{\lambda=1}^n \B{\lambda i}{\lambda j}
+ 2 \sum_{\lambda=1}^n \sum_{\eta < \lambda} \R{\eta i}{\lambda j}.
\label{sumrowB}
\eeq
Comparison of equation~(\ref{sumrowB}) with equations~(\ref{boxbox}) and 
(\ref{vproda}) reveals an interesting property of the matrix ${\cal B}$:
\beq
\sum_{\mbox{\small{column}} of {\cal B}} \mbox{Re}\left({\cal B}\right) = 
-\half \sum_{\lambda=1}^n |V_{\lambda i}|^2 |V_{\lambda j}|^2,
\eeq
where the sum is over a column of ${\cal B}$ specified by fixed $i$ and $j$.
We may in a
similar manner show that 
\beq
\sum_{\mbox{\small{row}} of {\cal B}} \mbox{Re}\left({\cal B}\right) = 
-\half \sum_{x=1}^n |V_{\alpha x}|^2|V_{\beta x}|^2,
\eeq
with constant $\alpha$ and $\beta$ specifying the row of ${\cal B}$ to be
summed.  Sums over rows are preferred to sums over columns phenomenologically 
when $n>3$, because complete rows are accessible whereas complete columns 
are not.

For $n\ge 3$, some $R$s will be positive while others are negative.  An upper
bound on the values of the positive $R$s may be established from the Schwarz
inequality for two-component complex ``vectors''
$\vec{V}_{\alpha}\equiv\left(\Vai, \V{\alpha j}\right)$ and
$\vec{V}_{\beta}\equiv\left(\V{\beta i},\Vbj\right)$:
\beq
\left|\Vai\Vbis+\V{\alpha j}\Vs{\beta j}\right|^2 \le 
  \left[\left|\Vai\right|^2 + \left|\V{\alpha j}\right|^2 \right]
  \left[\left|\V{\beta i}\right|^2 + \left|\Vbj\right|^2 \right].
\eeq
A bit of algebra gives the desired result:
\beq
\R{\alpha i}{\beta j} \le \half \left[ 
  \left|\Vai\right|^2 \left|\Vbj\right|^2 + 
  \left|\V{\alpha j}\right|^2 \left|\V{\beta i}\right|^2\right].
\eeq
Later we will actually derive an {\em equality} in three generations  
for $\R{\alpha i}{\beta j}$ 
in terms of the four $|V|$s appearing in this equation (found in
equation~(\ref{RVs})).

An alternative way to obtain constraints from unitarity is to start with the
definition of the boxes (\ref{boxdef}) and use the unitarity of the mixing
matrix:
\beq
\Baibj = \left(V_{\alpha i} \Vs{\alpha j} \right)
\left(V_{\beta j} \Vs{\beta i} \right) = 
\left(\delta_{ij} - \sum_{\eta \neq \alpha} V_{\eta i} \Vs{\eta j}\right)
\left(\delta_{ij} - \sum_{\lambda \neq \beta} V_{\lambda j}\Vs{\lambda
i}\right).
\label{profunit}
\eeq
After a bit of algebra, this becomes\footnote{
This relation also follows from equation~(\ref{urow}):
\bea
\sum_{\lambda \neq \beta} \sum_{\eta \neq \alpha} \B{\eta i}{\lambda j} &=& 
\sum_{\lambda \neq \beta} \left(\sum_{\eta=1}^n \B{\eta i}{\lambda j} - 
\B{\alpha i}{\lambda j}\right) = \sum_{\lambda \neq \beta} 
\left(|V_{\lambda i}|^2 \delta_{ij} - \B{\alpha i}{\lambda j}\right) \nonumber
\\
&=& \delta_{ij} \sum_{\lambda \neq \beta} |V_{\lambda i}|^2 - 
\left(\sum_{\lambda=1}^n \B{\alpha i}{\lambda j} - \Baibj\right) \\
&=& \delta_{ij} \left(1-|V_{\beta i}|^2\right) -|V_{\alpha i}|^2 \delta_{ij} 
+ \Baibj
\nonumber 
\eea 
}
\beq
\sum_{\eta \neq \alpha} \sum_{\lambda \neq \beta} \B{\eta i}{\lambda j}
=\Baibj - \delta_{ij} \left(-1+|V_{\alpha i}|^2 + |V_{\beta i}|^2\right).
\eeq

The number of unitarity constraints from equations~(\ref{urow}) and 
(\ref{ucol}), both real and imaginary, grows as $n^3$, while 
the number of ordered boxes grows as $n^4$, so additional relationships
between ordered boxes must exist to identify the $(n-1)^2$ independent $J$s
and $R$s.  
These additional identities will come from the definitions of the boxes, and the
resulting tautologies (\ref{vproda}) and (\ref{vprodi}), each of which grows in
number as
$n^6$, implying a high degree of redundancy.  Some of these
relationships are explored below in equations~(\ref{raa}) to
(\ref{realconstraint}).

The constraints (\ref{urow}) and (\ref{ucol}) 
hold independently for the real and imaginary parts of each sum.  
We will first explore the implications of these constraints for the imaginary
parts of boxes, before
turning to the implications of these constraints for the real parts.


\subsection{Unitarity Constraints on the Imaginary Parts of Boxes}


Because the right-hand sides of the equations
are manifestly real, as are terms on the left-hand side involving 
degenerate boxes, the sums of nondegenerate boxes in 
equations (\ref{urow}) and (\ref{ucol}) 
must be real.  This leads to imaginary constraints of the form\footnote{
The constraints (\ref{imagfixrow}) and (\ref{imagfixcol}) 
may be written exclusively in terms of ordered boxes as
\bea
\sum_{\eta < \lambda} \J{\eta i}{\lambda j} - 
\sum_{\eta > \lambda} \J{\lambda i}{\eta j} & = & 0, 
\mbox{\ \ \ and \ \ \ } \label{imagsumrow}\\
\sum_{y < x} \J{\alpha y}{\beta x} - 
\sum_{y > x} \J{\alpha x}{\beta y} & = & 0.
\label{imagsumcol}
\eea
While expressing everything in terms of ordered boxes is our goal, we will
find it more convenient to use the complete sums of equations~(\ref{imagfixrow})
and (\ref{imagfixcol}) in mathematical manipulations and switch to ordered boxes
at the end rather than to deal with the two separate terms in
equations~(\ref{imagsumrow}) and (\ref{imagsumcol}).
}
\bea
\sum_{\eta \neq \lambda} \J{\eta i}{\lambda j} & =& 0, 
\mbox{\ \ \ and \ \ \ } \label{imagfixrow}\\
\sum_{y \neq x} \J{\alpha y}{\beta x} &=& 0.
\label{imagfixcol}
\eea

In three generations, each sum in equation~(\ref{imagfixrow}) or
(\ref{imagfixcol}) contains only two terms, leading to the equality (up to a
sign) of two $J$s.  For example,
choosing $\lambda=1$ in three generations yields
\beq
\J{1j}{2i} = -\J{1j}{3i}, \mbox{\ \ \ }n=3,
\eeq
so
\beq
\J{11}{22} = -\J{11}{32}, \mbox{\ \ \ }n=3.
\eeq
We may continue in a similar manner and show that every $J$ is related to
${\cal J} \equiv \J{11}{22}$ \cite{Jarls1}, giving
\beq
\mbox{Im} ({\cal B}) = \left( \ba{ccc} 
{\cal J} &  {\cal J} & -{\cal J} \\
{\cal J} &  {\cal J} & -{\cal J} \\
-{\cal J} &  -{\cal J} & {\cal J}
\ea \right), \mbox{\ \ \ }n=3.
\eeq
One consequence of the equality up to a sign of all $J$s 
in three generations is that if any one $\Vai$ is zero, then all 
$\J{\alpha i}{\beta j}$ vanish and there can be no CP-violation.

For $n>3$ generations, the sums in
equations~(\ref{imagfixrow}) and (\ref{imagfixcol}) contain more than two terms,
so individual $J$s become equal to sums of other $J$s rather than
just equal to another individual $J$.  In four
generations, we know from counting phases in the mixing matrix, or more
rigorously from the work
in reference \cite{KS} that we 
should find three independent $J$s out of the $36$ ordered $J$s.
This reduction, unlike the reduction in three generations, is not achieved
solely by equations of the forms (\ref{imagfixrow}) and (\ref{imagfixcol}). 
No more than 27 independent constraints may be derived from those equations, so
at least 6 constraints on the imaginary parts of boxes in four generations will
come from the expressions derived below which relate real and imaginary 
parameters.


\subsection{Unitarity Constraints on the Real Parts of Boxes}


We now consider the real parts of the constraints (\ref{urow}) and (\ref{ucol}),

considering first the homogeneous constraints 
for which the Kronecker delta
is zero.  These relationships give the singly-degenerate boxes as sums of
ordered boxes:
\bea
|V_{\lambda i}|^2|V_{\lambda j}|^2 = 
\B{\lambda i}{\lambda j} &=& - \sum_{\eta \neq \lambda} \B{\eta i}{\lambda j}
=- \sum_{\eta \neq \lambda} \R{\eta i}{\lambda j},
\hspace{0.4 in} i \neq j, \mbox{\ \ and \ \ } 
\label{1degrow}\\
|V_{\alpha x}|^2|V_{\beta x}|^2 = 
\B{\alpha x}{\beta x} &=& - \sum_{y \neq x} \B{\alpha y}{\beta x} = 
- \sum_{y \neq x} \R{\alpha y}{\beta x} ,
\ssp \alpha \neq \beta.
\label{1degcol}
\eea
For three generations, each of the sums contains two terms, allowing us to
express the singly-degenerate boxes in terms of two nondegenerate boxes
measurable in neutrino appearance oscillation experiments.

For fixed $(i,j)$ in equation~(\ref{1degrow}), $\lambda$ can take $n$ 
possible values, implying $n$ constraint equations.  $N$
ordered nondegenerate boxes appear in these $n$ equations.  Thus, for $N\le n$,
which is true for $n\le 3$, 
the unitarity constraint (\ref{1degrow}) and (\ref{1degcol}) may be inverted to
find a nondegenerate box in terms of singly-degenerate boxes.  
Adding the sums in equation~(\ref{1degcol}) 
for $x=i$ and $x=j$, then subtracting the $x=k$ sum 
yields the desired expression:
\beq
\R{\alpha i}{\beta j} = -\half\left(
|V_{\alpha i}|^2|V_{\beta i}|^2 + |V_{\alpha j}|^2|V_{\beta j}|^2 - 
|V_{\alpha k}|^2|V_{\beta k}|^2\right), \mbox{\ \ \ }n=3,
\label{RKim1}
\eeq
where we have used the property that $R$s are real and therefore 
unchanged under mass (or flavor) index interchanges.
Many sources, such as reference~\cite{KimPev}, use these sums as the
coefficients of the oscillatory terms in three-flavor oscillation probabilities
to write
\bea
\label{KimPevsum}
\Pab &=& 2 \left( \sin^2 \Pt + \sin^2 \Pu - \sin^2 \Pe \right) 
  |V_{\alpha 1}|^2 |V_{\beta 1}|^2 \nonumber \\
&&+ \ 2 \left( \sin^2 \Pt - \sin^2 \Pu + \sin^2 \Pe \right)
  |V_{\alpha 2}|^2 |V_{\beta 2}|^2 \\
&&+ \ 2 \left( -\sin^2 \Pt + \sin^2 \Pu + \sin^2 \Pe \right)
  |V_{\alpha 3}|^2 |V_{\beta 3}|^2, \mbox{\ \ \ }n=3. \nonumber
\eea
Similar manipulation of equation~(\ref{1degcol}) gives an
alternate expression in term of the flavor triad $(\alpha, \beta, \gamma)$:
\beq
\R{\alpha i}{\beta j} = -\half \left(
|V_{\alpha i}|^2|V_{\alpha j}|^2 + |V_{\beta i}|^2|V_{\beta j}|^2 - 
|V_{\gamma i}|^2|V_{\gamma j}|^2 \right), \mbox{\ \ \ }n=3.
\label{RKim2}
\eeq

A simpler derivation of equations~(\ref{RKim1}) and (\ref{RKim2}) is available. 
Three-generation unitarity gives
\beq
\Vai\Vbis + \V{\alpha j}\Vs{\beta j} = - \V{\alpha k}\Vs{\beta k}.
\eeq
Squaring both sides of this equation then leads directly to
equation~(\ref{RKim1}).  Generalizing this derivation to $n>3$, we see why a
single nondegenerate box cannot be expressed as a sum of degenerate boxes for
$n>3$, but we also obtain some new relations among boxes.  For example, with
$n=4$, we may square both sides of 
\beq
\V{\alpha 1} \Vs{\beta 1} +  \V{\alpha 2} \Vs{\beta 2} = 
-\V{\alpha 1} \Vs{\beta 1} - \V{\alpha 1} \Vs{\beta 1}
\eeq
to obtain
\beq
\R{\alpha 1}{\beta 2} - \R{\alpha 3}{\beta 4} = \half \left[ 
\left|\V{\alpha 3}\right|^2 \left|\V{\beta 3}\right|^2 + 
\left|\V{\alpha 4}\right|^2 \left|\V{\beta 4}\right|^2 - 
\left|\V{\alpha 1}\right|^2 \left|\V{\beta 1}\right|^2 - 
\left|\V{\alpha 2}\right|^2 \left|\V{\beta 2}\right|^2  \right], \mbox{\ \ \
}n=4.
\eeq

It has been shown that knowledge of four $|V|$s completely specifies the
three-generation mixing matrix, provided no more than two $|V|$s are
taken from the same row or same column \cite{HnJ}.  Therefore,
it may prove useful to use three-generation unitarity to re-write
equation~(\ref{RKim2}) in terms of just four $|V|$s.  The result is
\beq
\R{\alpha i}{\beta j} = \half \left[1-\left|\Vai\right|^2 - 
\left|\V{\alpha j}\right|^2 - \left|\V{\beta i}\right|^2 - \left|\Vbj\right|^2 +
\left|\Vai\right|^2\left|\Vbj\right|^2 + 
\left|\V{\alpha j}\right|^2 \left|\V{\beta i}\right|^2 \right], \mbox{\ \ \
}n=3.
\label{RVs}
\eeq
This has an interesting appearance, for it expresses the real part of the box
Re$\left[\Vai\Vajs\Vbj\Vbis\right]$ in terms of the magnitudes of the four
complex $V$s which define the box.  

The unitarity constraints (\ref{1degrow}) and (\ref{1degcol}) 
greatly simplify our expressions for a
doubly-degenerate box $\B{\alpha i}{\alpha i} = |\V{\alpha i}|^2$:
\bea
\B{\alpha i}{\alpha i} & = & 
\frac{\B{\alpha i}{\alpha x} \B{\alpha i}{\alpha y}} {\B{\alpha x}{\alpha y}} = 
\frac{\left(- \sum_{\eta \neq \alpha} \R{\alpha i}{\eta x} \right)
\left(- \sum_{\lambda \neq \alpha} \R{\alpha i}{\lambda y} \right)}
{\left(- \sum_{\tau \neq \alpha} \R{\alpha x}{\tau y} \right)},
\ssp x \neq y \neq i, \mbox{\ \ and \ \ } 
\label{2degxy} \\
{\B{\alpha i}{\alpha i}} & = &
\frac{\B{\alpha i}{\lambda i} \B{\alpha i}{\eta i}}{\B{\lambda i}{\eta i}} =
\frac{\left(- \sum_{x \neq i} \R{\alpha x}{\lambda i}  \right)
\left(- \sum_{y \neq i} \R{\alpha y}{\eta i}  \right)}
{\left( - \sum_{z \neq i} \R{\lambda z}{\eta i}  \right)},
\hspace{0.45 in} \lambda \neq \eta \neq \alpha,
\label{2degetlam}
\eea
where the first equalities are due to equations~(\ref{vproda}) and
(\ref{vprodi}).  Applying equation~(\ref{2degxy}) to three generations, one
finds that doubly-degenerate boxes are expressible 
in terms of the real parts of 
six ordered boxes, rather than the nine complex boxes 
used in equations (\ref{vfour}) and (\ref{vfour2}).  For example,
\beq
|V_{11}|^4 = \B{11}{11} = 
-\frac{\left(\R{11}{22}+\R{11}{32}\right) \left(\R{11}{23}+\R{11}{33}\right)}
{\R{12}{23}+\R{12}{33}}, \mbox{\ \ \ }n=3.
\label{B1111}
\eeq
When considering $n>3$, each sum has more terms, but all terms in the 
numerators in equations
(\ref{2degxy}) and (\ref{2degetlam}) always contain $R$s to the second order,
while the denominator terms contain only the first order of $R$s.  Thus these
expressions will be much more manageable than equations (\ref{vfour}) and
(\ref{vfour2}) which exhibit the fifth order of boxes in the numerator and the
fourth order in the denominator.

Summing equation~(\ref{1degrow}) over $j\neq i$ yields another 
expression for $|V_{\lambda i}|^2$ in terms of nondegenerate boxes:
\beq
|V_{\lambda i}|^2 \sum_{j\neq i} |V_{\lambda j}|^2 = 
|V_{\lambda i}|^2 \left(1-|V_{\lambda i}|^2\right) = 
-\sum_{j\neq i} \sum_{\eta \neq \lambda} \R{\eta i}{\lambda j}.
\eeq
The same result is obtained by summing equation~(\ref{urow}) over all $j$.  
The explicit solution or the above equation, valid for any number of
generations, has a two-fold ambiguity:
\beq
|V_{\lambda i}|^2 = \half\left[
1 \pm \sqrt{1+4\sum_{j\neq i} \sum_{\eta \neq \lambda} \R{\eta i}{\lambda
j}}\right].
\label{profeq}
\eeq
When the double sum is small, an approximation to the exact equation
(\ref{profeq}) is
\beq
|V_{\lambda i}|^2 \approx \left(
- \sum_{j\neq i} \sum_{\eta \neq \lambda} \R{\eta i}{\lambda j}, \ \ 
1+\sum_{j\neq i} \sum_{\eta \neq \lambda} \R{\eta i}{\lambda j}\right).
\eeq
For three generations, this approximation 
yields $|V_{\lambda i}|^2$ as a linear equation of
four nondegenerate boxes.  For example,
\bea
|V_{12}|^2 &=& \half \left[ 1 \pm
\sqrt{1+4\left(\R{11}{22}+\R{11}{32}+\R{12}{23}+\R{12}{33}\right)}\right]
\nonumber \\
& \approx & \left[-\left(\R{11}{22}+\R{11}{32}+\R{12}{23}+\R{12}{33}\right), \ \
1+\left(\R{11}{22}+\R{11}{32}+\R{12}{23}+\R{12}{33}\right)\right], \mbox{\ \ \
}n=3. \nonumber 
\eea


\subsection{Unitarity Relationships Relating Real Parts of Boxes with Imaginary
Parts of Boxes}


We may use the homogeneous unitarity conditions 
(\ref{1degrow}) and (\ref{1degcol}), along with the
tautologies (\ref{vproda}) and (\ref{vprodi}) to obtain constraints between
nondegenerate boxes, thereby reducing the number of real degrees of freedom. 
Recognizing that the tautology (\ref{vproda}) 
gives
\bea
\B{\alpha i}{\alpha j} & = & 
\mbox{Re} \left(\frac{\Bs{\alpha i}{\eta j} \B{\alpha i}{\lambda j}}
{\B{\eta i}{\lambda j}} \right) \nonumber \\
& = & 
{
{\frac{\R{\alpha i}{\eta j} \R{\alpha i}{\lambda j} \R{\eta i}{\lambda j} +
\J{\alpha i}{\eta j} \J{\alpha i}{\lambda j} \R{\eta i}{\lambda j} - 
\J{\alpha i}{\eta j} \R{\alpha i}{\lambda j} \J{\eta i}{\lambda j} +
\R{\alpha i}{\eta j} \J{\alpha i}{\lambda j} \J{\eta i}{\lambda j}}
{\left(\R{\eta i}{\lambda j}\right)^2 + \left(\J{\eta i}{\lambda j}\right)^2}}},
\label{raa} 
\eea
the unitarity constraint (\ref{1degrow}) 
becomes
{\small{
\beq
\R{\alpha i}{\eta j} \R{\alpha i}{\lambda j} \R{\eta i}{\lambda j} +
\J{\alpha i}{\eta j} \J{\alpha i}{\lambda j} \R{\eta i}{\lambda j} - 
\J{\alpha i}{\eta j} \R{\alpha i}{\lambda j} \J{\eta i}{\lambda j} +
\R{\alpha i}{\eta j} \J{\alpha i}{\lambda j} \J{\eta i}{\lambda j}
= - \left(\left(\R{\eta i}{\lambda j}\right)^2 + 
 \left(\J{\eta i}{\lambda j}\right)^2 \right)
 \sum_{\tau \neq \alpha} \R{\alpha i}{\tau j},  
\label{realdegaa}
\eeq
}}
with the usual inequalities $\eta \neq \lambda \neq \alpha$, $i \neq j$
satisfied.  A similar treatment may be applied to equation ~(\ref{vprodi}) to
arrive at the relation
{\small{
\beq
\R{\alpha i}{\beta x} \R{\alpha i}{\beta y} \R{\alpha x}{\beta y} +
\J{\alpha i}{\beta x} \J{\alpha i}{\beta y} \R{\alpha x}{\beta y} - 
\J{\alpha i}{\beta x} \R{\alpha i}{\beta y} \J{\alpha x}{\beta y} +
\R{\alpha i}{\beta x} \J{\alpha i}{\beta y} \J{\alpha x}{\beta y} 
= - \left(\left(\R{\alpha x}{\beta y}\right)^2 + 
 \left(\J{\alpha x}{\beta y}\right)^2 \right)
 \sum_{z \neq i} \R{\alpha i}{\beta z}, 
\label{realdegii}
\eeq
}}
with $\alpha \neq \beta$, and $x \neq y \neq i$. 
These constraints interrelate imaginary and real parts of boxes through
unitarity for any number of generations.  

We may infer constraints  similar to equations~(\ref{realdegaa}) and
(\ref{realdegii}) 
by setting the imaginary part of a singly-degenerate box to zero:
\bea
\mbox{Im} \left(\B{\alpha i}{\alpha j}\right) = 
\mbox{Im} \left( \frac{\Bs{\alpha i}{\eta j} \B{\alpha i}{\lambda j}}
{\B{\eta i}{\lambda j}} \right)=  \hspace{3 in}&& \label{taut1} \\
  \frac{\R{\alpha i}{\eta j} \J{\alpha i}{\lambda j} \R{\eta i}{\lambda j}
     - \J{\alpha i}{\eta j} \R{\alpha i}{\lambda j} \R{\eta i}{\lambda j}
     - \R{\alpha i}{\eta j} \R{\alpha i}{\lambda j} \J{\eta i}{\lambda j}
     - \J{\alpha i}{\eta j} \J{\alpha i}{\lambda j} \J{\eta i}{\lambda j}}
 {\R{\eta i}{\lambda j}^2 + \J{\eta i}{\lambda j}^2} = 0, && \nonumber
\eea
($\eta \neq \lambda \neq \alpha$, $i \neq j$), 
and
\bea 
\mbox{Im} \left(\B{\alpha i}{\beta i}\right) = 
\mbox{Im} \left( \frac{\Bs{\alpha i}{\beta x} \B{\alpha i}{\beta y}}
{\B{\alpha x}{\beta y}} \right) = \hspace{3 in} && \\
\frac{\R{\alpha i}{\beta x} \J{\alpha i}{\beta y} \R{\alpha x}{\beta y}
     - \J{\alpha i}{\beta x} \R{\alpha i}{\beta y} \R{\alpha x}{\beta y}
     - \R{\alpha i}{\beta x} \R{\alpha i}{\beta y} \J{\alpha x}{\beta y}
     - \J{\alpha i}{\beta x} \J{\alpha i}{\beta y} \J{\alpha x}{\beta y}}
 {\R{\alpha x}{\beta y}^2 + \J{\alpha x}{\beta y}^2} = 0, && \nonumber
\eea
($\alpha \neq \beta$, and $x \neq y \neq i$), so
\beq
\R{\alpha i}{\eta j} \J{\alpha i}{\lambda j} \R{\eta i}{\lambda j}
  - \J{\alpha i}{\eta j} \R{\alpha i}{\lambda j} \R{\eta i}{\lambda j}
  - \R{\alpha i}{\eta j} \R{\alpha i}{\lambda j} \J{\eta i}{\lambda j}
  - \J{\alpha i}{\eta j} \J{\alpha i}{\lambda j} \J{\eta i}{\lambda j} =0,
\label{homog1}
\eeq
($\eta \neq \lambda \neq \alpha$, $i \neq j$), and
\beq
\R{\alpha i}{\beta x} \J{\alpha i}{\beta y} \R{\alpha x}{\beta y}
  - \J{\alpha i}{\beta x} \R{\alpha i}{\beta y} \R{\alpha x}{\beta y}
  - \R{\alpha i}{\beta x} \R{\alpha i}{\beta y} \J{\alpha x}{\beta y}
  - \J{\alpha i}{\beta x} \J{\alpha i}{\beta y} \J{\alpha x}{\beta y} = 0.
\label{homog2}
\eeq
($\alpha \neq \beta$, and $x \neq y \neq i$).  
With a bit of index switching, these constraints lead to the isolation of a
single $R$:
\beq
\R{\alpha i}{\beta j} = 
  \frac{\J{\alpha i}{\beta j} \R{\alpha i}{\lambda j} \R{\beta i}{\lambda j}
    + \J{\alpha i}{\beta j} \J{\alpha i}{\lambda j} \J{\beta i}{\lambda j}}
  {\J{\alpha i}{\lambda j} \R{\beta i}{\lambda j}  
    - \R{\alpha i}{\lambda j} \J{\beta i}{\lambda j}} = 
   \frac{\J{\alpha i}{\beta j} \R{\alpha i}{\beta y} \R{\alpha j}{\beta y}
    + \J{\alpha i}{\beta j} \J{\alpha i}{\beta y} \J{\alpha j}{\beta y}}
  {\J{\alpha i}{\beta y} \R{\alpha j}{\beta y}
    - \R{\alpha i}{\beta y} \J{\alpha j}{\beta y}},
\label{realconstraint}
\eeq
with $\eta \neq \lambda \neq \alpha$, $i \neq j$ in the first expression, and 
$\alpha \neq \beta$, and $x \neq y \neq i$ in the second.


\subsection{Indirect Measurement of CP-Violation}


Suppose CP is conserved.  Then $\J{\alpha i}{\beta j}=0$ for all index choices. 
The implication for equations~(\ref{realdegaa}) and (\ref{realdegii}) are
\beq
\R{\alpha i}{\eta j} \R{\alpha i}{\lambda j} + \R{\eta i}{\lambda j}
\sum_{\tau \neq \alpha} \R{\alpha i}{\tau j} = 0,
\hspace{0.5 in} (\eta \neq \lambda \neq \alpha, \mbox{\ and \ } i\neq j), 
\eeq
and
\beq
\R{\alpha i}{\beta x}\R{\alpha i}{\beta y} + \R{\alpha x}{\beta y} 
\sum_{z\neq i} \R{\alpha i}{\beta z} = 0, 
\hspace{0.5 in}(x\neq y \neq z, \mbox{\ and \ } \alpha \neq \beta).
\eeq
If either of these relations is
violated, then so is CP.  An attempt to determine the validity of these
relations may be facilitated by substituting 
$-\half\sum_{\alpha=1}^n \left|\V{\alpha i}\right|^2\left|\V{\alpha j}\right|^2$
for the sum over $\tau$ and
$-\half\sum_{i=1}^n \left|\V{\alpha i}\right|^2 \left|\V{\beta i}\right|^2$
for the sum over $z$, as given in equations~(\ref{1degrow}) and (\ref{1degcol}).

For three generations, $\tau$ must equal $\lambda$ in 
equation~(\ref{realdegaa}), $\J{\alpha i}{\beta j}= -\J{\alpha i}{\lambda j} = 
\J{\beta i}{\lambda j}={\cal J}$ by equation~(\ref{imagfixrow}), and
equation (\ref{realdegaa}) may be solved for ${\cal J}^2$ directly:
\beq
{\cal J}^2  = 
\R{\alpha i}{\beta j} \R{\beta i}{\lambda j} + 
\R{\alpha i}{\beta j} \R{\alpha i}{\lambda j} + 
\R{\alpha i}{\lambda j} \R{\beta i}{\lambda j}, \mbox{\ \ \ }n=3,
\label{CPone} 
\eeq
which expresses ${\cal J}$ in terms of $3$ $R$s summed in pairs.
Similarly, in equation~(\ref{realdegii}), $z=y$, and $\J{\alpha i}{\beta j} = 
- \J{\alpha i}{\beta y} = 
\J{\alpha j}{\beta y}$ by equation (\ref{imagfixcol}), leading to another
expression for ${\cal J}^2$:
\beq
{\cal J}^2  =  \R{\alpha i}{\beta j} \R{\alpha
j}{\beta y} + 
\R{\alpha i}{\beta j} \R{\alpha i}{\beta y} + 
\R{\alpha j}{\beta y} \R{\alpha i}{\beta y}, \mbox{\ \ \ }n=3.
\label{CPtwo}
\eeq
These relationships between the $R$s and $J$s in three generations
exhibit a simple parameter symmetry:  ${\cal J}^2$ in
equation~(\ref{CPone}) equals $\R{\alpha i}{\beta j} \R{\lambda i}{\beta j}+ $ 
terms cyclic in $(\alpha, \beta, \lambda)$; in equation~(\ref{CPtwo}), 
it equals $\R{\alpha i}{\beta j} \R{\alpha y}{\beta j}+$ terms cyclic in
$(i, j, y)$.  Equations~(\ref{CPone}) and (\ref{CPtwo}) say that the three real
elements in any row or column of the matrix ${\cal B}$ may be summed in their
three pairwise products to yield the CP-violating invariant ${\cal J}^2$.  Note
that if CP is conserved and ${\cal J}$ is zero, then equations~(\ref{CPone}) and
(\ref{CPtwo}) also tell us that all three $R$s in any column or row cannot have
the same sign.

If CP is violated, ${\cal J} \neq 0$, so the combination of real
parts on the right-hand sides of equations~(\ref{CPone}) and (\ref{CPtwo}),
measurable with CP-conserving averaged neutrino
oscillations, cannot be zero.
Thus, even if CP violating asymmetries are not
directly observable in an experiment, the effects of CP violation may be seen
through the relationships among the real parts of different boxes!  A related
observation has been made for the quark sector by Hamzaoui in
reference~\cite{HnJ}.

For three generations, unitarity enforces the simple
relations~(\ref{imagfixrow})
and (\ref{imagfixcol}) among the $J$s in equation~(\ref{realconstraint}), which
then reproduces the expressions (\ref{CPone}) and (\ref{CPtwo}).
For $n>3$, however,
equation (\ref{realconstraint}) is different from the latter equations,
since the sums $\sum_{\tau \neq \acb}$ and $\sum_{z \neq i,j}$ in
equations~(\ref{realdegaa}) and (\ref{realdegii}) ensure that
boxes with all $n$ flavor or mass indices enter into the equations replacing 
(\ref{CPone}) or (\ref{CPtwo}), respectively.  Equation (\ref{realconstraint}) 
involves only three such indices for any $n$, 
so it will be different from the unitarity constraints in higher
generations, providing additional identities among the real parts of boxes.


\subsection{Inhomogeneous Unitarity Constraints}


The homogeneous unitarity constraints cannot provide the desired
normalization of the $V_{\alpha i}$ or the boxes; we need the
inhomogeneous unitarity constraints arising when the Kronecker delta in
equations~(\ref{urow}) and (\ref{ucol})
is not zero.  We now develop these inhomogeneous 
constraints\footnote{
In terms of mixing-matrix elements, equations~(\ref{2degsuma}) and
(\ref{2degsumi}) are trivial.  For example, equation~(\ref{2degsuma}) is
\beq
|\Vai|^4 + \sum_{\eta \neq \alpha} |\Vai|^2|V_{\eta i}|^2 = |\Vai|^2,
\mbox{\ \ \ or \ \ }
\sum_{\eta=1}^n |V_{\eta i}|^2 = 1.
\eeq
}
which are functions of degenerate boxes and therefore purely real:
\bea
\B{\alpha i}{\alpha i} + \sum_{\eta \neq \alpha} \B{\alpha i}{\eta i} &= &
\sqrt{\B{\alpha i}{\alpha i}}, \mbox{\ \ and \ \ }
\label{2degsuma} \\
\B{\alpha i}{\alpha i} + \sum_{z \neq i} \B{\alpha i}{\alpha z} &=& 
\sqrt{\B{\alpha i}{\alpha i}}.
\label{2degsumi}
\eea
Parenthetically, we note by 
comparing equations~(\ref{2degsuma}) and (\ref{2degsumi}) that a sum
over mass-degenerate boxes equals a sum over flavor-degenerate boxes:
\beq
\sum_{\eta \neq \alpha} \B{\alpha i}{\eta i} =
 \sum_{z \neq i} \B{\alpha i}{\alpha z}.
\eeq

Equations~(\ref{2degsuma}) and (\ref{2degsumi}) 
can be rewritten strictly in terms of nondegenerate boxes 
by using the homogeneous unitarity constraints
(\ref{1degrow}) and (\ref{1degcol}). We find
\bea
&& \hspace{-2.5 cm} 
\frac{\left(- \sum_{\lambda \neq \alpha} \R{\alpha i}{\lambda x} \right)
\left(- \sum_{\sigma \neq \alpha} \R{\alpha i}{\sigma y} \right)}
{\left(- \sum_{\tau \neq \alpha} \R{\alpha x}{\tau y} \right)} \nonumber \\
&& - \ \sqrt{\frac{\left(- \sum_{\lambda \neq \alpha} 
\R{\alpha i}{\lambda x} \right)
\left(- \sum_{\sigma \neq \alpha} \R{\alpha i}{\sigma y} \right)}
{\left(- \sum_{\tau \neq \alpha} \R{\alpha x}{\tau y} \right)}} 
+ \sum_{\eta \neq \alpha} \left(- \sum_{z \neq i} \R{\alpha z}{\eta i} \right) 
= 0,
\eea
with $x \neq y \neq i$, and
\bea
&& \hspace{-2.5 cm} 
\frac{\left(- \sum_{x \neq i} \R{\alpha x}{\lambda i}  \right)
\left(- \sum_{y \neq i} \R{\alpha y}{\eta i}  \right)}
{\left( - \sum_{t \neq i} \R{\lambda t}{\eta i}  \right)} \nonumber \\
&& - \ \sqrt{\frac{\left(- \sum_{x \neq i} \R{\alpha x}{\lambda i}  \right)
\left(- \sum_{y \neq i} \R{\alpha y}{\eta i}  \right)}
{\left( - \sum_{t \neq i} \R{\lambda t}{\eta i}  \right)}}
+ \sum_{z \neq i} \left(- \sum_{\eta \neq \alpha} \R{\alpha z}{\eta i} \right)
= 0,
\eea
with $\lambda \neq \eta \neq \alpha$.  Note that these inhomogeneous unitarity
constraints do not involve the $J$s.

Isolating the square root, squaring the equation, and multiplying 
through by the resulting denominator, we get quartic equations, each relating
$n(n-1)$ $R$s:
\beq
\ba{l}
\left(\sum_{\lambda \neq \alpha} \R{\alpha i}{\lambda x} \right)
\left(\sum_{\sigma \neq \alpha} \R{\alpha i}{\sigma y} \right)
\left(\sum_{\tau \neq \alpha} \R{\alpha x}{\tau y} \right)
\left[1+2\left(\sum_{\eta \neq \alpha} \sum_{z \neq i} \R{\alpha z}{\eta i}
\right)\right] +
\\
\hspace{0.2 cm} 
\left(\sum_{\lambda \neq \alpha} \R{\alpha i}{\lambda x} \right)^2
\left(\sum_{\sigma \neq \alpha} \R{\alpha i}{\sigma y} \right)^2 + 
\left(\sum_{\tau \neq \alpha} \R{\alpha x}{\tau y} \right)^2 
\left(\sum_{\eta \neq \alpha} \sum_{z \neq i} \R{\alpha z}{\eta i} \right)^2
=0,
\ea
\label{Vfouraa}
\eeq
($x \neq y \neq i$), and
\beq
\ba{l}
\left(\sum_{x \neq i} \R{\alpha x}{\lambda i}  \right)
\left(\sum_{y \neq i} \R{\alpha y}{\eta i}  \right)
\left(\sum_{t \neq i} \R{\lambda t}{\eta i}  \right)
\left[1+2\left(\sum_{z \neq i}\sum_{\eta \neq \alpha} \R{\alpha z}{\eta i} 
\right) \right] +
\\
\hspace{0.2 cm} 
\left(\sum_{x \neq i} \R{\alpha x}{\lambda i}  \right)^2
\left(\sum_{y \neq i} \R{\alpha y}{\eta i}  \right)^2 +
\left(\sum_{t \neq i} \R{\lambda t}{\eta i}  \right)^2
\left(\sum_{z \neq i}\sum_{\eta \neq \alpha} \R{\alpha z}{\eta i} \right)^2
=0
\ea
\label{Vfourii}
\eeq
for $\lambda \neq \eta \neq \alpha$.  
In three generations, each of the sums in equations~(\ref{Vfouraa}) and
(\ref{Vfourii}) has only two terms, which is not so formidable.  For example,
for $\alpha =2$ and  $i = 1$, we have
\beq
\ba{l}
\left(\R{11}{22}+\R{21}{32}\right)\left(\R{11}{23}+\R{21}{33}\right)
\left(\R{12}{23}+\R{22}{33}\right) 
\left[1+2\left(\R{11}{22}+\R{21}{32}+\R{11}{23}+\R{21}{33}\right)\right] +
\\
\hspace{0.2 cm} 
\left(\R{11}{22}+\R{21}{32}\right)^2 \left(\R{11}{23}+\R{21}{33}\right)^2 + 
\left(\R{11}{22}+\R{21}{32}+\R{11}{23}+\R{21}{33}\right)^2
\left(\R{12}{23}+\R{22}{33}\right)^2 = 0, \mbox{\ \ \ }n=3,
\ea
\label{V11fouraa}
\eeq
and
\beq
\ba{l}
\left(\R{11}{22}+\R{11}{23}\right)\left(\R{21}{32}+\R{21}{33}\right)
\left(\R{11}{32}+\R{11}{33}\right) 
\left[1+2\left(\R{11}{22}+\R{11}{23}+\R{21}{32}+\R{21}{33}\right)\right] +
\\
\hspace{0.2 cm} 
\left(\R{11}{22}+\R{11}{23}\right)^2 \left(\R{21}{32}+\R{21}{33}\right)^2 + 
\left(\R{11}{22}+\R{11}{23}+\R{21}{32}+\R{21}{33}\right)^2
\left(\R{11}{32}+\R{11}{33}\right)^2 = 0, \mbox{\ \ \ }n=3.
\ea
\label{V11fourii}
\eeq

\section{Reduction to a basis}


\subsection{General Algorithm}


As with the constraints on the imaginary parts of boxes, many of the
constraints (\ref{raa}) to (\ref{Vfourii})
are redundant, but an independent set may be used to reduce the number of box
parameters to a basis.  The homogeneous constraints~(\ref{realdegaa}),
(\ref{realdegii}), and 
(\ref{realconstraint}) are much simpler than the
inhomogeneous constraints~(\ref{Vfouraa}) and (\ref{Vfourii}), 
so it is advantageous 
to use as many homogeneous constraints as possible to construct the basis. 
Still, some inhomogeneous constraints must be invoked if the boxes and matrix 
elements of V are to be normalized.  

Notice that with only fixed $i$ and $j$
mass indices and a flavor sum over the entire $(i,j)$ column of ${\cal B}$, 
in equation~(\ref{realdegaa}), and only fixed $\alpha$ and
$\beta$ flavor indices and a mass sum over the entire $(\acb)$ row of ${\cal B}$
in equation~(\ref{realdegii}), these
equations express relationships among all the boxes 
within single columns and rows, respectively, of
the matrix ${\cal B}$ defined in equation~(\ref{boxbox}).
For each choice for the set of indices,
equations~(\ref{realdegaa}) and (\ref{realdegii}) may be used to express one $R$
or $J$ in terms of the others.  This suggests a general algorithm for
constructing a basis set of boxes.  One $R$ from each row and column may be
taken as dependent using the homogeneous unitarity constraints~(\ref{realdegaa})
and (\ref{realdegii}), or (\ref{realconstraint}).  This leaves
$(n-1)^2$ $R$s and $q$ $J$s in the independent set, where $q$ is the number of
$J$s which are still independent after the application of the 
homogeneous constraints equations~(\ref{imagfixrow}) and (\ref{imagfixcol}). 
The $n\times n$ mixing matrix is parameterized by $(n-1)^2$ parameters, so we
seek a basis set of $(n-1)^2$ box parameters.  The inhomogeneous unitarity
equations, which we will address shortly, must supply the remaining constraints.
The number of inhomogeneous constraints to be used is apparently equal to $q$. 
We have seen that with three flavors this is just one.  The end result is a
basis of $(n-1)^2$ $R$s and $J$s with no more than $n-1$ $R$s coming from any
row
or any column.  If one restricts the bases and counting to accessible boxes, the
algorithm may differ for $n>3$.  We do not pursue this here.

\subsection{Three Generations}

We provide here an example of a basis construction for three generations
obtained by substituting in the unitarity equations derived above.
Rearranging the three-generation equations~(\ref{CPone}) and (\ref{CPtwo}) 
yields expressions for one $R$ in terms of two other $R$s and ${\cal J}$:
\bea
\R{\alpha i}{\beta j} & = & \frac{{\cal J}^2 - 
\R{\alpha i}{\lambda j}\R{\beta i}{\lambda j}}
{\R{\beta i}{\lambda j}+\R{\alpha i}{\lambda j}} 
\label{R3genaa} \\
& = & \frac{{\cal J}^2 - 
\R{\alpha j}{\beta y}\R{\alpha i}{\beta y}}
{\R{\alpha j}{\beta y}+\R{\alpha i}{\beta y}}
\label{R3genii} 
\eea
We may then
eliminate $\R{12}{23}$ by either equation~(\ref{R3genaa}) or
equation~(\ref{R3genii})
\beq
\R{12}{23} = \frac{\R{12}{33}\R{22}{33} - {\cal J}^2}{-\R{22}{33} - \R{12}{33}} 
           = \frac{\R{11}{22}\R{11}{23} - {\cal J}^2}{-\R{11}{23} - \R{11}{22}}.
\label{1223out}
\eeq
We may similarly eliminate $\R{11}{32}$ and $\R{21}{33}$
\beq
\R{11}{32} = \frac{\R{11}{22}\R{21}{32} - {\cal J}^2}{-\R{21}{32} - \R{11}{22}}
           = \frac{\R{11}{33}\R{12}{33} - {\cal J}^2}{-\R{12}{33} - \R{11}{33}},
\label{1132out}
\eeq
and
\beq
\R{21}{33} = \frac{\R{11}{23}\R{11}{33} - {\cal J}^2}{-\R{11}{33} - \R{11}{23}}
           = \frac{\R{21}{32}\R{22}{33} - {\cal J}^2}{-\R{22}{33} -\R{21}{32}}.
\label{2133out}
\eeq
$\R{12}{33}$ may be eliminated from equation (\ref{1223out}), and $\R{11}{33}$
may be eliminated from equation (\ref{2133out}):
\bea
\R{12}{33} &=& \frac{\R{22}{33} \R{11}{22} \R{11}{23} 
  + {\cal J}^2 \left( \R{11}{23} + \R{11}{22} - \R{22}{33} \right)}
  {\R{22}{33} \R{11}{23} + \R{22}{33} \R{11}{22} - \R{11}{22} \R{11}{23} +
  {\cal J}^2}, \mbox{ \ \ and \ \ } \label{1233out} \\
\nonumber \\
\R{11}{33} &=& \frac{\R{22}{33} \R{11}{23} \R{21}{32} 
  + {\cal J}^2 \left( \R{21}{32} + \R{22}{33} - \R{11}{23} \right)}
  {\R{11}{23} \R{21}{32} + \R{22}{33} \R{11}{23} - \R{21}{32} \R{22}{33} +
  {\cal J}^2}. \label{1133out}
\eea
Equation~(\ref{1132out}) will not provide an additional constraint; it is
redundant to the other two.  As expected, we must turn to the inhomogeneous
constraints to eliminate the last degree of freedom.    
We will here choose the constraint (\ref{V11fouraa}), 
since its expression in terms of our four remaining
boxes is the least complicated.  We may substitute the second equality
from equation~(\ref{2133out}) for $\R{21}{33}$ and the second equality of
equation~(\ref{1223out}) for $\R{12}{23}$ into equation~(\ref{V11fouraa}).

Multiplying through to place all of the terms in the numerator, we are left with
a constraint which contains the five parameters $\R{11}{22}$, 
$\R{11}{23}$, $\R{21}{32}$, $\R{22}{33}$, and ${\cal J}^2$:
\bea
&&\hspace{-0.5 cm} 
0=\left(\R{11}{22}+\R{11}{23}\right)^2 \left(\R{11}{22}+\R{21}{32}\right)^2 
\left[-\R{21}{32} \R{22}{33} + 
\R{11}{23}\left(\R{21}{32} + \R{22}{33}\right) + {\cal J}^2\right]^2 
\nonumber \\
&& + \left[ \left(\R{21}{32}\right)^2 + \left(\R{11}{23} +
+ \R{11}{22} \right) 
\left( \R{21}{32} + \R{22}{33} \right) + {\cal J}^2 \right]^2
\nonumber \\
&& \ssp \times \left[\R{11}{23} \R{22}{33} + \R{11}{22}
\left(-\R{11}{23}+\R{22}{33} \right)
{\cal J}^2 \right]^2 
\label{quartic} \\
&&- \ \left(\R{11}{22}+\R{21}{32}\right) \left(\R{11}{22}+\R{11}{23}\right)
\left[\R{11}{22} \left(\R{11}{23}-\R{22}{33}\right) - \R{11}{23} \R{22}{33}
- {\cal J}^2\right] \nonumber \\
&& \ssp \times \left[\R{21}{32} + 2 \R{11}{22} \R{21}{32} + 
2 \left(\R{21}{32}\right)^2 + 2 \R{21}{32} \R{11}{23} \right. \nonumber \\
&& \left.  + \ \R{22}{33} +
2 \R{11}{22} \R{22}{33} + 2 \R{11}{23} \R{22}{33} + 2 {\cal J}^2\right]
\left[\R{11}{23}\left(\R{21}{32}+\R{22}{33}\right)-\R{21}{32}\R{22}{33} +
{\cal J}^2\right].
\nonumber
\eea
This equation is quartic in all five parameters.  We may eliminate any one 
by either algebraic or numeric means, leaving us with the desired 
four parameters as the basis.


\section{summary}


Neutrino physics has entered a golden age of research.  New experiments all over
the globe promise an unequaled amount of data from the sun, the atmosphere,
accelerators, supernovae, and other cosmic sources.  The latest data suggests
that more than three neutrino flavors may participate in neutrino oscillations 
\cite{BWW}.  Analyzing such refined data
requires a consistent, model-independent approach which may be easily applied,
and easily extended
to higher generations.  Our work presented here offers such an approach.  One
result which we view as particularly noteworthy is that high-statistics data on
{\em averaged} oscillations is sufficient to determine the conservation or
non-conservation of CP in the lepton mixing matrix; this indirect test of CP is
a consequence of the unitarity of the mixing matrix.

\section*{Acknowledgements}

This work was supported in part by the U.S. Department of Energy, Division of
High Energy Physics, under Grant No. DE-F605-85ER40226.


\appendix  

\section{Graphical Representation of Box Relations}
\label{graphapp}

\begin{figure}[htb]
\vsp
\begin{centering}
\begin{picture}(310,90)(-5,5)
\thicklines
\multiput(0,90)(30,0){3}{\circle*{4}}
\multiput(0,60)(30,0){3}{\circle*{4}}
\multiput(0,30)(30,0){3}{\circle*{4}}
\put(0,87){\vector(0,-1){24}}
\put(30,63){\vector(0,1){24}}
\put(24,10){(a)}
\multiput(120,90)(30,0){3}{\circle*{4}}
\multiput(120,60)(30,0){3}{\circle*{4}}
\multiput(120,30)(30,0){3}{\circle*{4}}
\put(123,90){\vector(1,0){24}}
\put(147,60){\vector(-1,0){24}}
\put(144,10){(b)}
\multiput(240,90)(30,0){3}{\circle*{4}}
\multiput(240,60)(30,0){3}{\circle*{4}}
\multiput(240,30)(30,0){3}{\circle*{4}}
\put(240,63){\vector(0,1){24}}
\put(270,87){\vector(0,-1){24}}
\put(264,10){(c)}
\end{picture}
\caption{The graphical representation for (a) $\B{11}{22}$, \ (b)
$\left(\B{11}{22}\right)^{-1}$, \ and (c) $\Bs{11}{22}$.
\label{graphfig}}
\end{centering}
\end{figure}
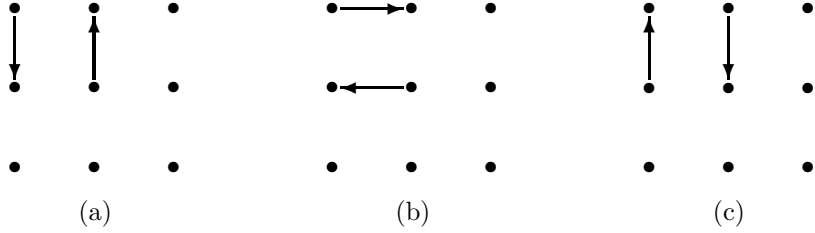

Many of the relationships between boxes developed in Section~\ref{boxVsec} can
be
derived using a graphical representation.  In this method,
boxes in the numerator of a product are represented by two vertical lines; boxes
in
the denominator are represented by two horizontal lines.  Lines exit the
locations
of the matrix elements which are not complex-conjugated 
and enter the locations of the complex-conjugated
matrix elements.  For example, the box $\B{11}{22}=V_{11} V^*_{12} V_{22}
V^*_{21}$ is represented by a
vertical line pointing from $V_{11}$ to $V_{21}$ and a vertical line pointing
from $V_{22}$ to $V_{12}$, as shown in Figure~\ref{graphfig}a.  The inverse box
$\frac{1}{\B{11}{22}} = \left(\B{11}{22}\right)^{-1}$ is represented by a
horizontal line pointing from $V_{11}$ to $V_{12}$ and one pointing from 
$V_{22}$ to $V_{21}$, as shown in Figure~\ref{graphfig}b.  The
complex-conjugated box $\Bs{11}{22}$ is equal to $\B{12}{21}$, so one just
reverses the arrows to complex conjugate a box, 
as shown in Figure~\ref{graphfig}c.

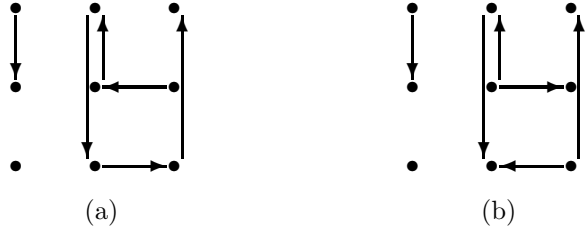
\begin{figure}[thb]
\begin{centering}
\begin{picture}(220,100)(-5,5)
\thicklines
\multiput(0,90)(30,0){3}{\circle*{4}}
\multiput(0,60)(30,0){3}{\circle*{4}}
\multiput(0,30)(30,0){3}{\circle*{4}}
\put(0,87){\vector(0,-1){24}}
\put(27,87){\vector(0,-1){54}}
\put(33,63){\vector(0,1){24}}
\put(63,33){\vector(0,1){54}}
\put(57,60){\vector(-1,0){24}}
\put(33,30){\vector(1,0){24}}
\put(26,10){(a)}
\multiput(150,90)(30,0){3}{\circle*{4}}
\multiput(150,60)(30,0){3}{\circle*{4}}
\multiput(150,30)(30,0){3}{\circle*{4}}
\put(150,87){\vector(0,-1){24}}
\put(177,87){\vector(0,-1){54}}
\put(183,63){\vector(0,1){24}}
\put(213,33){\vector(0,1){54}}
\put(207,30){\vector(-1,0){24}}
\put(183,60){\vector(1,0){24}}
\put(176,10){(b)}
\end{picture}
\caption{The graphical representation for the products 
a) $\B{11}{22} \B{12}{33} \left(\B{23}{32}\right)^{-1}$, and 
b)$\B{11}{22} \B{12}{33} \left(\B{22}{33}\right)^{-1}$.
\label{cancelfig}}
\end{centering}
\end{figure}

To multiply boxes together graphically, one merely draws the lines corresponding
to each factor on the same grid.  Horizontal arrows entering or leaving a point
cancel out vertical arrows entering or leaving, respectively, that point. 
Uncanceled arrows entering a point signify the survival of the
complex-conjugated matrix element associated with that point.  Those leaving a
point signify the survival of the ordinary matrix element of that point. 
Figure~\ref{cancelfig}a represents the product 
$\B{11}{22} \B{12}{33} \left(\B{23}{32}\right)^{-1}$.  The horizontal arrows do
not cancel vertical arrows at any point, so no simplification may occur.  The
upward arrow at $V_{22}$ represents that element in the numerator.  The
horizontal arrow represents the element $V^*_{22}$ in the denominator, which
does not cancel.  Counting off the arrows at each vertex, we find the expression
\beq
\B{11}{22} \B{12}{33} \left(\B{23}{32}\right)^{-1} = 
\frac{V_{11} V_{12} V^*_{12} V^*_{13} V^*_{21} V_{22} V^*_{32} V_{33}}
{V^*_{22} V_{23} V_{32} V^*_{33}},
\eeq
which agrees with the definitions of boxes.  

Figure~\ref{cancelfig}b represents
$\B{11}{22} \B{12}{33} \left(\B{22}{33}\right)^{-1}$, a product in which
some canceling does occur.  Picking out the uncanceled arrows at each vertex, we
are left with
\beq
\B{11}{22} \B{12}{33} \left(\B{22}{33}\right)^{-1} = 
\frac{V_{11} V_{12} V^*_{12} V^*_{13} V^*_{21}}{V^*_{23}}.
\eeq

The graphical method is a powerful tool for finding relationships among boxes.
For example, trying to obtain the equations~(\ref{vfour}) and (\ref{vfour2}) for

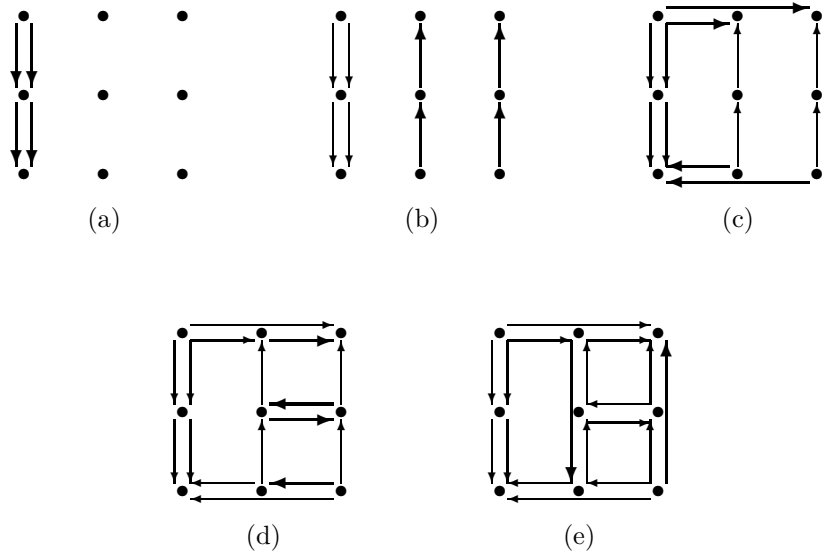
\begin{figure}[t]
\begin{centering}
\begin{picture}(310,220)(-5,5)
\thicklines
\multiput(0,210)(30,0){3}{\circle*{4}}
\multiput(0,180)(30,0){3}{\circle*{4}}
\multiput(0,150)(30,0){3}{\circle*{4}}
\put(-3,207){\vector(0,-1){24}}
\put(3,207){\vector(0,-1){24}}
\put(-3,177){\vector(0,-1){24}}
\put(3,177){\vector(0,-1){24}}
\put(24,130){(a)}
\multiput(120,210)(30,0){3}{\circle*{4}}
\multiput(120,180)(30,0){3}{\circle*{4}}
\multiput(120,150)(30,0){3}{\circle*{4}}
\thinlines
\put(117,207){\vector(0,-1){24}}
\put(123,207){\vector(0,-1){24}}
\put(117,177){\vector(0,-1){24}}
\put(123,177){\vector(0,-1){24}}
\thicklines
\put(150,183){\vector(0,1){24}}
\put(180,183){\vector(0,1){24}}
\put(150,153){\vector(0,1){24}}
\put(180,153){\vector(0,1){24}}
\put(144,130){(b)}
\multiput(240,210)(30,0){3}{\circle*{4}}
\multiput(240,180)(30,0){3}{\circle*{4}}
\multiput(240,150)(30,0){3}{\circle*{4}}
\thinlines
\put(237,207){\vector(0,-1){24}}
\put(243,207){\vector(0,-1){24}}
\put(237,177){\vector(0,-1){24}}
\put(243,177){\vector(0,-1){24}}
\put(270,183){\vector(0,1){24}}
\put(300,183){\vector(0,1){24}}
\put(270,153){\vector(0,1){24}}
\put(300,153){\vector(0,1){24}}
\thicklines
\put(243,213){\vector(1,0){54}}
\put(243,207){\vector(1,0){24}}
\put(297,147){\vector(-1,0){54}}
\put(267,153){\vector(-1,0){24}}
\put(264,130){(c)}
\multiput(60,90)(30,0){3}{\circle*{4}}
\multiput(60,60)(30,0){3}{\circle*{4}}
\multiput(60,30)(30,0){3}{\circle*{4}}
\thinlines
\put(57,87){\vector(0,-1){24}}
\put(63,87){\vector(0,-1){24}}
\put(57,57){\vector(0,-1){24}}
\put(63,57){\vector(0,-1){24}}
\put(90,63){\vector(0,1){24}}
\put(120,63){\vector(0,1){24}}
\put(90,33){\vector(0,1){24}}
\put(120,33){\vector(0,1){24}}
\put(63,93){\vector(1,0){54}}
\put(63,87){\vector(1,0){24}}
\put(117,27){\vector(-1,0){54}}
\put(87,33){\vector(-1,0){24}}
\thicklines
\put(93,87){\vector(1,0){24}}
\put(117,63){\vector(-1,0){24}}
\put(93,57){\vector(1,0){24}}
\put(117,33){\vector(-1,0){24}}
\put(84,10){(d)}
\multiput(180,90)(30,0){3}{\circle*{4}}
\multiput(180,60)(30,0){3}{\circle*{4}}
\multiput(180,30)(30,0){3}{\circle*{4}}
\thinlines
\put(177,87){\vector(0,-1){24}}
\put(183,87){\vector(0,-1){24}}
\put(177,57){\vector(0,-1){24}}
\put(183,57){\vector(0,-1){24}}
\put(213,63){\vector(0,1){24}}
\put(237,63){\vector(0,1){24}}
\put(213,33){\vector(0,1){24}}
\put(237,33){\vector(0,1){24}}
\put(183,93){\vector(1,0){54}}
\put(183,87){\vector(1,0){24}}
\put(237,27){\vector(-1,0){54}}
\put(207,33){\vector(-1,0){24}}
\put(213,87){\vector(1,0){24}}
\put(237,63){\vector(-1,0){24}}
\put(213,56){\vector(1,0){24}}
\put(237,33){\vector(-1,0){24}}
\thicklines
\put(207,87){\vector(0,-1){54}}
\put(243,33){\vector(0,1){54}}
\put(204,10){(e)}
\end{picture}
\caption{The steps to obtaining $|V_{21}|^4$ as a function of ordered,
non-degenerate boxes.  The additions in each step are designated by the thicker
arrows.
\label{V21fig}}
\end{centering}
\end{figure}

$\B{\alpha i}{\alpha i} = |V_{\alpha i}|^4$ of
Section~\ref{boxVsec} without graphs involved quite a few false starts.  
Using the graphical method, we need only find a series of arrows which
cancel for every point except $(\alpha, i)$ and leave two incoming and two
outgoing vertical arrows at that point.  Consider $|V_{21}|^4$ as an example. 
We choose to draw all of the arrows involving $V_{21}$ pointing downward, as
shown in Figure~\ref{V21fig}a.  These arrows must be part of boxes, so in
Figure~\ref{V21fig}b we add the arrows to finish those boxes.  
Next we draw two horizontal boxes in Figure~\ref{V21fig}c to
cancel the extra arrows in the first column of the matrix.  This still leaves
$V_{22}$ and $V_{23}$ with two sets of uncanceled arrows apiece.  In
Figure~\ref{V21fig}d we draw two more horizontal boxes to compensate.  This adds
arrows to our previously clean $V_{12}$, $V_{13}$, $V_{32}$, and $V_{33}$.
Drawing the final vertical box in Figure~\ref{V21fig}e cancels those.  Recapping
what we have done, we see that step (b) completes $\B{11}{22}$, $\B{11}{23}$,
$\B{21}{32}$, and $\B{21}{33}$ in the numerator.  Step (c) divides by
$\B{11}{32}$ and $\B{11}{33}$, and step (d) divides by $\B{12}{23}$ and
$\B{22}{33}$.  Step (e) multiplies by $\B{12}{33}$, leaving only the point
$V_{21}$ with uncanceled arrows.  It has two vertical arrows coming in and two
leaving, so our graph represents the equation
$$
|V_{21}|^4 = \frac{\B{21}{32}\B{11}{22}\B{21}{33}\B{11}{23}\B{12}{33}}
{\B{11}{32}\B{11}{33}\B{22}{33}\B{12}{23}}
\eqno{(\ref{V21four})}
$$
of Section~\ref{boxVsec}.
Other examples of this representation are included in that section.

\end{document}